\documentclass[prb,twocolumn,preprintnumbers,amsmath,amssymb]{revtex4-2}
\usepackage{graphicx}% Include figure files
\usepackage{dcolumn}% Align table columns on decimal point
\usepackage{amsmath,amsbsy,amssymb,scalefnt}
\usepackage{bm}% bold math
\usepackage{epstopdf}
\usepackage[T1]{fontenc}
\usepackage{amsfonts}
\usepackage{booktabs}
\usepackage{colortbl}
\usepackage{textcomp}
\usepackage{mathtools}
\usepackage{xcolor}
\usepackage{hyperref}
\usepackage{physics}
\hypersetup{ pdfnewwindow=true, colorlinks=true, linkcolor=blue, anchorcolor=blue, citecolor=blue, filecolor=blue, menucolor=blue, urlcolor=blue}

\begin{document}
\title{Dissipationless transport by design in ultrathin magnetic topological insulator films}

\author{Amir Sabzalipour}
\affiliation{COMMIT, Department of Physics, University of Antwerp, Groenenborgerlaan 171, B-2020 Antwerp, Belgium}
\author{Mohammad Shafiei}
\affiliation{COMMIT, Department of Physics, University of Antwerp, Groenenborgerlaan 171, B-2020 Antwerp, Belgium}

\author{Milorad V. Milo\v{s}evi\'c}
\email{milorad.milosevic@uantwerpen.be}
\affiliation{COMMIT, Department of Physics, University of Antwerp, Groenenborgerlaan 171, B-2020 Antwerp, Belgium}

\date{\today}

\begin{abstract}
Magnetic topological insulators (MTIs) are among the prominent platforms for the next generation of high-speed and low-power spintronic devices. However, unlike their non-magnetic counterparts, where the surface spin-momentum locking prevents electrons from being scattered by non-magnetic impurities and results in a dissipationless electronic flow, magnetic impurities in MTIs cause dissipation by exerting magnetic torque on the electron spin. Decreasing this resistance is desired to reduce energy consumption and optimize performance of MTIs in envisaged applications. Here we reveal how electronic backscattering can be suppressed in a MTI thin film by external magnetic and/or electronic stimuli, to yield an entirely dissipationless spin-polarized charge transport. Our findings thus present an effective route to preserve spin coherence and enhance spin-current functionality in magnetic topological materials, suggesting design strategies for magneto-electronic and spintronic devices with strongly reduced energy consumption.
\end{abstract}

\maketitle

\section{Introduction}
Magnetic topological insulators (MTIs) have emerged in the last decade as an attractive class of quantum materials, providing a versatile platform to explore rich fundamental physics and enable development of next-generation spintronic devices~\cite{tokura2019magnetic,fan2016electric,che2020strongly}. Topological insulators (TIs) that are intrinsically magnetic do exist, such as MnBi$_2$Te$_4$~\cite{wang2021intrinsic}, but magnetization can also be introduced into TIs in a tailored manner, e.g. by proximity to magnetic layers~\cite{bhattacharyya2021recent}, or by controlled magnetic doping~\cite{chang2013experimental}. These MTIs exhibit diverse spin-dependent phenomena of high research interest~\cite{tokura2019magnetic}, such as spin-transfer torques~\cite{fischer2016spin}, efficient spin-charge interconversion~\cite{zhang2016conversion}, topological antiferromagnetic spintronics~\cite{vsmejkal2018topological}, the rare formation of magnetic skyrmions in topological systems~\cite{tokura2019magnetic}, to name a few.

%By introducing magnetization into TIs, through proximity to magnetic layers~\cite{bhattacharyya2021recent}, controlled magnetic doping~\cite{chang2013experimental}, or intrinsic magnetic compounds such as MnBi$_2$Te$_4$~\cite{wang2021intrinsic}, one can unlock a host of exotic spin-dependent phenomena~\cite{tokura2019magnetic}. These include spin-transfer torques~\cite{fischer2016spin}, efficient spin-charge interconversion~\cite{zhang2016conversion}, topological antiferromagnetic spintronics~\cite{vsmejkal2018topological}, and the formation of magnetic skyrmions in topological systems~\cite{tokura2019magnetic}.%

MTIs are particularly appealing for spintronic applications due to their ability to generate spin-polarized edge or surface currents under an applied bias, as a consequence of spin-momentum locking inherent to topological surface states~\cite{flatte2017voltage}. The resulting spin accumulation at the boundaries depends both on current direction and magnetization orientation, enabling robust spin filtering and the generation of spin-current~\cite{wu2014topological,he2019topological}. Furthermore, this spin accumulation can exert spin-transfer torques on adjacent magnetic layers~\cite{fan2016spintronics}, offering a route for efficient, low-power control of magnetic states. When combined with their high spin-charge conversion efficiency and chiral edge states that support topologically protected spin textures~\cite{yu2010quantized}, MTIs provide a broad platform for dissipationless spin transport and stand out as promising candidates for future spin-based logic and memory technologies ~\cite{li2014electrical,mellnik2014spin,wang2023room,binda2023large,wu2019room,he2019topological}.

%Among the various approaches to realize magnetic TIs, magnetic doping, such as chromium incorporation into (Bi/Sb)$_2$(Se/Te)$_3$~\cite{chang2013experimental}, offers precise control over the magnitude and direction of magnetization~\cite{haazen2012ferromagnetism}. This control is critical for designing tunable device functionalities and can be engineered during material growth or modified by post-synthesis. In Cr-doped TIs, ferromagnetism can emerge via bulk van Vleck mechanisms or via surface carrier-mediated exchange at low temperatures~\cite{kou2013interplay}. However, magnetic doping also introduces significant challenges: the distribution of magnetic impurities disrupts the coherence of topological surface states, leading to impurity scattering and disorder that restore dissipation such as Joule heating~\cite{tokura2019magnetic}. Consequently, such impurities disturb the inherently dissipationless nature of topological transport, posing a significant challenge for practical applications that rely on robust, low-energy electronic conduction ~\cite{yan2024rules}.

Among the various approaches to realize MTIs, magnetic doping, such as chromium incorporation into (Bi/Sb)$_2$(Se/Te)$_3$~\cite{chang2013experimental}, offers precise control over the net magnitude and direction of magnetization~\cite{haazen2012ferromagnetism}. This control is critical for designing tunable device functionalities and can be engineered during material growth or modified by post-synthesis. In Cr-doped TIs, the long-range magnetism can be supported by bulk van Vleck mechanisms or via surface carrier-mediated exchange at low temperatures~\cite{kou2013interplay}. 

However, while magnetic doping facilitates the manipulation of magnetic properties, it simultaneously introduces magnetic impurities that perturb the topological surface states. Due to the inherent spin-momentum locking in TIs, where electron spin orientation is rigidly tied to its momentum, these magnetic impurities induce spin-flip scattering that directly translates into momentum scattering, thereby causing energy dissipation~\cite{pop2010energy,vassighi2006thermal}. This dissipation, manifested as impurity scattering, disorder, and Joule heating~\cite{tokura2019magnetic}, undermines the dissipationless nature of topological transport and poses a significant challenge for practical applications that rely on robust, low-energy electronic conduction~\cite{yan2024rules}. Therefore, a deep understanding and mitigation of dissipation mechanisms in MTI thin films are essential for advancing their practical applications in energy-efficient electronics.

%In topological insulators (TIs), strong spin-orbit coupling leads to spin-momentum locking, where an electron's spin orientation is rigidly tied to its momentum. This coupling ensures that any perturbation affecting the spin directly influences the momentum, and vice versa. While spin-momentum locking underpins the topological protection of surface states, it also makes the system highly sensitive to magnetic perturbations. Magnetic impurities can induce spin-flip scattering, which, due to spin-momentum locking, translates into momentum scattering, leading to energy dissipation. These effects are detrimental for nanoscale spintronic applications, where thermal management and energy efficiency are critical~\cite{pop2010energy,vassighi2006thermal}. Therefore, understanding and mitigating dissipation mechanisms in magnetic TI thin films is essential for harnessing their full potential in ultralow-power electronic systems.

In this work, we address this critical challenge of energy dissipation in magnetically doped ultrathin TIs by providing a framework for achieving dissipationless longitudinal charge and spin transport. Focusing on magnetically doped ultrathin Bi$_2$Se$_3$ as a representative example, we investigate how the interplay between magnetic configuration, gate-tunability, and topological gapped surface states can be exploited to fully suppress electron scattering. Our analysis begins with an effective low-energy Hamiltonian for the massive Dirac surface states in momentum space and employs a semiclassical Boltzmann approach to compute the transport properties. To provide a closer description of realistic systems and validate our findings, we further perform real-space simulations based on a tight-binding model and calculate the conductance/resistance using the Landauer-B\"uttiker formalism.

The close agreement between these two methods confirms that electron scattering in magnetically doped TI films can be strongly and even entirely suppressed through electrostatic gating and/or magnetization tilting. Moreover, we reveal that the spin polarization of the charge current, which is typically weakened as a result of hybridization effects and spin decoherence from magnetic impurities, can be effectively restored. By applying a gate voltage to induce asymmetry between the top and bottom surfaces of the film, we reduce the hybridization of the surface states, thereby enhancing spin-momentum locking. Additionally, tuning the magnetization orientation of the magnetic dopants with respect to the spin texture of the conducting states further reinforces spin coherence. Collectively, these strategies not only deepen our understanding of dissipation mechanisms in MTIs but also provide a viable route toward ultralow-power spintronic devices with robust spin-polarized transport.

The paper is organized as follows. In Sec.~\ref{system and Hamiltonian}, we detail the system under investigation, and discuss the effective Hamiltonian of its surface states. Our main results are laid out in Sec.~\ref{main results}, starting from the scattering analysis to demonstrate the successful manipulation of back-scattering by magnetization orientation and/or gating, and the consequently broadly tuned resistance and dissipation in the system. The latter is then validated in Sec.~\ref{validation real-space} by real-space transport calculations within the tight-binding approach, where more of the realistic features of the system are taken into account. Sec.~\ref{conclusions} offers a summary of our findings and conclusions.

\section{Description of the system and the governing Hamiltonian}\label{system and Hamiltonian}
\begin{figure}[t]
    \centering
    \includegraphics[width=0.9\linewidth]{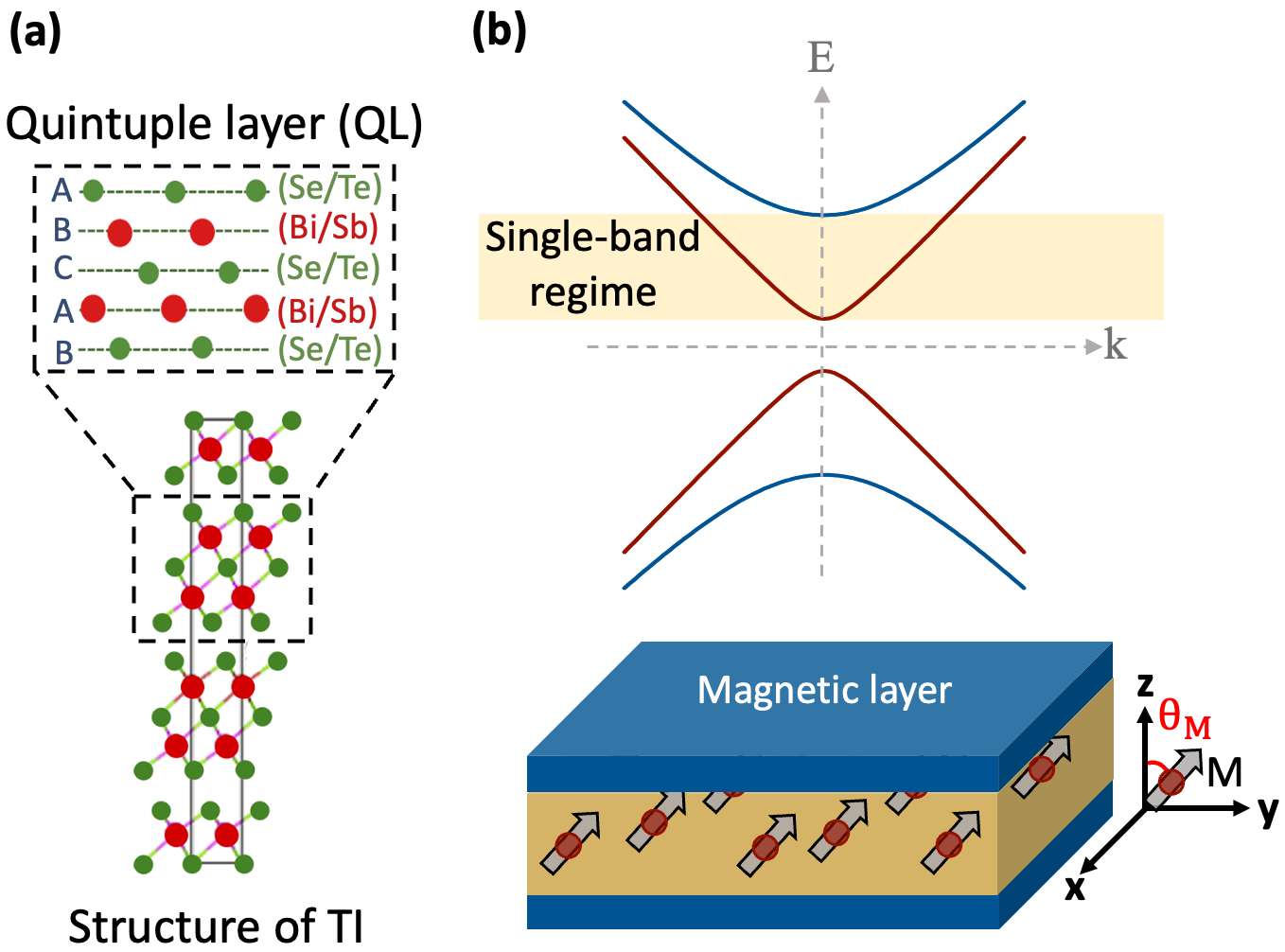}
    \caption{(a) Crystal structure of the Bi$_2$Se$_3$ TI family, consisting of quintuple layers arranged in a rhombohedral lattice. (b) Schematic illustration of the system under study, a magnetically doped ultrathin TI film sandwiched between two magnetic layers, and its single-band regime. The proximity-induced exchange field lifts the degeneracy of the conduction subbands. The Fermi level is then tuned into the conduction regime, where only the lower surface conduction band is occupied.}
    \label{fig:fig1}
\end{figure}

We consider an ultrathin, magnetically doped Bi$_2$Se$_3$ film, where intersurface coupling, enhanced by quantum confinement, gives rise to a rich landscape of spin-charge phenomena. This material family is known for its strong spin-orbit coupling, topologically nontrivial band structure, and layered rhombohedral crystal structure~\cite{zhang2009topological}, with each unit cell comprising two Bi and three Se atoms, forming a quintuple layer (QL)~\cite{zhang2009topological,zhang2010crossover} as shown in Fig.~\ref{fig:fig1}(a). In ultrathin films, typically a few QLs thick, the top and bottom surface states are no longer independent but overlap and hybridize, resulting in the opening of a hybridization gap at the Dirac point~\cite{zhang2010crossover,shafiei2024tuning}. This regime is particularly relevant for device miniaturization and for tuning the electronic and topological properties via electrostatic gating or magnetic proximity effects.

To describe the low-energy surface physics of this system, we employ a momentum-space effective Hamiltonian as~\cite{shan2010effective}:

\begin{equation}\label{hamil_momen}
\begin{split}
&H(\mathbf{k}) = \, (E_{0}-Dk^{2})\sigma_0 \otimes \tau_0 + \hbar v_F (  k_y \sigma_x -  k_x\sigma_y)\otimes\tau_z \\
& + (\Delta_h - Bk^2)\sigma_0 \otimes \tau_x + V_{SIA}\,\sigma_0 \otimes \tau_z+ \textbf{M}\cdot\sigma \otimes \tau_0,
\end{split}
\end{equation}
where $\mathbf{k} = (k_x, k_y)$ is the in-plane wave vector, ${\sigma_i}$ and ${\tau_i}$ are Pauli matrices acting on spin and layer (top/bottom surface) degrees of freedom, respectively. $v_F$ is the Fermi velocity of electrons and $\Delta_h$ is the hybridization gap at the $\Gamma$ point. The terms $Dk^2$ and $Bk^2$ are quadratic terms in momentum, associated with particle-hole asymmetry and the hybridization gap at finite $\mathbf{k}$, respectively. $E_0$ results in a uniform energy shift and does not affect the band structure of the system. $\textbf{M}$ denotes the exchange interaction induced by the intrinsic magnetization of the system. The term $V_{\text{SIA}}\,\sigma_0 \otimes \tau_z$ captures the effect of structural inversion asymmetry (SIA), which arises in ultrathin films due to the presence of a substrate~\cite{zhang2010crossover}. This asymmetry, modeled as an effective potential along the film normal, lifts the top-bottom surface degeneracy and induces Rashba-type spin splitting, affecting surface-state dispersion and spin transport. The parameters of this Hamiltonian are thickness-dependent and have been extracted in literature by fitting experimental data using analytical solutions. We adopt these values from Ref.~\cite{zhang2010crossover}, as summarized in Table~\ref{tab:params}. %It should be noted that, although the value of V$_{SIA}$ is taken from experimental data on ultrathin films, the actual V$_{SIA}$ will vary depending on the specific experimental implementation of our proposed setup. The total potential in the system upon applying an external gate voltage is then considered as the sum of the contributions from the structural inversion asymmetry and the gate voltage, i.e., V$_{total}$ = V$_{SIA}$ + V$_g$.

The system is considered as a dilute magnetically doped ultrathin TI film. Magnetic doping is introduced through localized magnetic moments throughout the crystal lattice, which collectively yield a net magnetization vector $\mathbf{M}$. When the magnetization is aligned perpendicular to the film plane, the resulting exchange coupling opens a magnetic gap in the surface Dirac cones, characterized by an energy scale $\Delta_M$. In the present case, given that the system is in the dilute doping regime, the induced $\Delta_M$ is small compared to the intrinsic gap of the system and can therefore be neglected.

\begin{table}[b]
\centering
\begin{tabular}{|c|c|c|c|c|c|}
\bottomrule
\bottomrule
%\rowcolors{1}{gray!20}{white} % Alternate row coloring
\textbf{\,\,\,\,\,\,\,\,\,\,\,\,\,\,\,\,\,\,\,\,} & \textbf{2QL} & \textbf{3QL} & \textbf{4QL} & \textbf{5QL} & \textbf{$\geq$6QL} \\
%\bottomrule
\midrule
$E_0$ (eV) & -0.470 & -0.407 & -0.363 & -0.345 & -0.324 \\
$D$ (eV $\textup{\AA}^2)$ & -14.4 & -9.7 & -8.0 & -15.3 & -13.0 \\
$v_F$ (eV $\textup{\AA}$) & 3.10 & 3.17 & 2.95 & 2.99 & 2.98\\
$\Delta_h$ (eV) & 0.252 & 0.138 & 0.070 & 0.041 & 0 \\
$B$ (eV $\textup{\AA}^2)$ & 21.8 & 18 & 10 & 5 & 0 \\
V$_{SIA}$ (eV) & 0 & 0.038 & 0.053 & 0.057 & 0.068 \\
\bottomrule
\bottomrule
\end{tabular}
\caption{Thickness-dependent parameters $\Delta_0$, $\Delta_1$, and $v_F$ for ultrathin Bi$_2$Se$_3$ films (of thickness 2-6QL). Values derived from experimental data in Ref.~\onlinecite{zhang2010crossover}.}
\label{tab:params}
\end{table}

To allow for controllable tilting of the intrinsic magnetization, we consider a magnetically doped ultrathin TI film sandwiched between two magnetic layers, as illustrated in Fig.~\ref{fig:fig1}(b). In such a heterostructure, the magnetization direction of the magnetic layers ensures the alignment of magnetic dopants in the TI along a desired tilt angle $\theta_M$, defined in the $yz$-plane with respect to the $z$-axis (cf. Fig.~\ref{fig:fig1}(b)). As we will show, such a control is essential for the tunability of transport properties of this system.

Another important role of the magnetic layers in this heterostructure is to ensure that TI is in the single-band regime, i.e. the regime where only one Dirac surface band intersects the Fermi level, as illustrated in Fig.~\ref{fig:fig1}(b). To efficiently control the energy loss during the transport of Fermi electrons in magnetically doped TI films, it is essential to suppress transitions between gapped surface states. These transitions are highly anisotropic, and their contribution to scattering can significantly degrade the spin coherence and transport efficiency. A critical step to mitigate this dissipation is to eliminate one of the two conductive surface channels from participating in transport. The proximity effect of the magnetic layers is modeled as a Zeeman exchange coupling, which enters the Hamiltonian as $\Delta_{t/b} \sigma_z$, where $\Delta_t$ and $\Delta_b$ represent the exchange coupling strengths on the top and bottom surfaces, respectively. Due to the broken time-reversal symmetry introduced by proximity magnetization of two different magnetic layers \footnote{In Ref.~\cite{hou2020axion}, first-principles calculations have demonstrated that sandwiching a TI film between CrI$_3$ and MnBi$_2$Se$_4$ induces distinctly different exchange fields (approximately 2.9~meV and 26.9~meV, respectively) on the top and bottom surfaces of the TI.}, the degeneracy between the conduction subbands is lifted, leading to significant energy separation of the conduction bands. The magnitude of this separation is determined by $|\Delta_t - \Delta_b|$, and in this work we consider an energy splitting of 35~meV, which can realistically be induced via the proximity effect by coupling the TI to appropriate magnetic insulators~\cite{li2022large,eremeev2018new}. This separation allows the upper conduction band to be excluded from transport by tuning the chemical potential to lie within a single conduction band. 

While such a single-band regime offers theoretical and practical advantages, it is not without experimental challenges. Precise tuning of the Fermi level to lie within a single surface band requires advanced control via chemical doping or electrostatic gating. However, these techniques are rapidly improving, following the progress in material synthesis, nanofabrication, and gating technologies.

%Although our primary focus is on the single-band regime, our real-space transport calculations also encompass the two-band regime, in which both conduction bands contribute to transport. Although the presence of interband scattering in this regime introduces additional complexity and modifies the transport behavior compared to the single-band case, we find that the system’s transport characteristics remain tunable. In particular, our results indicate that dissipation associated with surface states can still be effectively reduced through a combination of magnetization tilting and electrostatic gating. These findings underscore the robustness and generality of the proposed control mechanisms, extending their applicability beyond the idealized single-band scenario.

\section{Dissipationless transport in ultrathin magnetic TI film}\label{main results}
In this section, we delve into the conditions under which dissipationless transport emerges in magnetically doped ultrathin topological insulator films. We systematically explore how magnetization orientation and external gate voltage influence electron scattering and, in turn, the longitudinal resistivity of the system. We begin by examining the impact of magnetization tilting, demonstrating that alignment of magnetic moments along particular directions can significantly reduce intraband scattering. Next, we explore the effect of electrostatic gating on the transport characteristics, highlighting its role in tuning the electronic structure and enhancing robustness against scattering. Finally, we investigate the combined influence of magnetization and gating, identifying regimes where their interplay leads to the maximal suppression of scattering channels and the onset of (nearly) dissipationless conduction.

\subsection{Effect of magnetization orientation}
We first examine how the orientation of magnetic impurities, aligned via proximity to ferromagnetic insulators and/or by tilted external magnetic field, influences transport in magnetically doped TI films. The electron-impurity interaction is modeled as a localized exchange potential $V_{sc}(\boldsymbol{r}) = J_0 \delta(\boldsymbol{r} - \boldsymbol{R}_{im})\, \boldsymbol{S}_{im} \cdot \boldsymbol{S}_e$, where all impurity spins are aligned at a tilt angle $\theta_M$ in the $y$–$z$ plane: $\boldsymbol{S}_{im} = S_{im}(0, \sin\theta_M, \cos\theta_M)$. The impact of this angular alignment on scattering processes and resistivity is explored in the following.

\begin{figure}[b]
    \centering
    \includegraphics[width=0.8\linewidth]{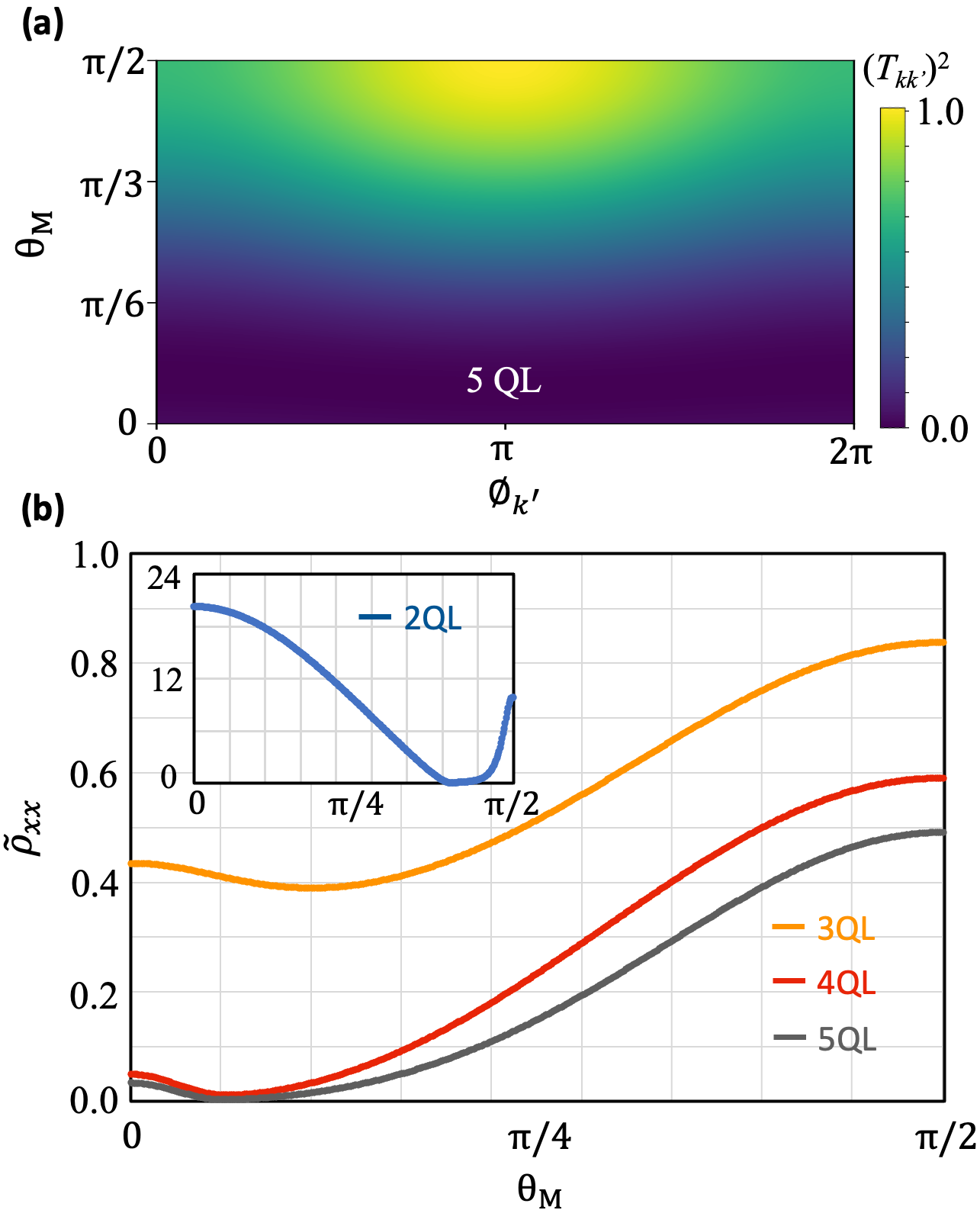}
    \caption{(a) Scattering probability $T_{\boldsymbol{k} \boldsymbol{k}'}^2$ as a function of magnetization angle $\theta_M$ and outgoing momentum angle $\phi_{\boldsymbol{k}'}$ for a 5QL-thick Bi$_2$Se$_3$ film under an in-plane electric field $\boldsymbol{E} = E \hat{x}$ with incoming electron momentum $\phi_{\boldsymbol{k}} = \pi$. (b) Longitudinal resistivity $\rho_{xx}$ as a function of magnetization tilt angle $\theta_M$, for magnetically doped Bi$_2$Se$_3$ films with thicknesses of 2-5~QL, in absence of gating ($V_{g}=0$). The non-monotonic behavior highlights the tunability of resistive transport via magnetic orientation, with thickness-dependent resistivity minima emerging at intermediate angles.}
    \label{fig:fig2}
\end{figure}
To quantify the effect of magnetic impurities on longitudinal transport, we employ the $T$-matrix formalism within the first Born approximation (see, Ref.~\onlinecite{kohn1957quantum, vyborny2009semiclassical}). The scattering amplitude is given by $T(\boldsymbol{k}, \boldsymbol{k}') = \langle \boldsymbol{k} | V_{sc}(\boldsymbol{r}) | \boldsymbol{k}' \rangle$, where $|\boldsymbol{k} \rangle$ and $|\boldsymbol{k}' \rangle$ are eigenstates of the unperturbed Hamiltonian [Eq.~\eqref{hamil_momen}]. This framework allows us to evaluate spin-dependent scattering as a function of magnetization orientation, providing insights into tunable transport behavior in magnetically doped TI films.

The square of the scattering matrix element $T_{\boldsymbol{k} \boldsymbol{k}'}^2$, is derived analytically and depends explicitly on the relative angles of electron momentum and magnetization orientation. The angular dependence results in highly anisotropic scattering profiles, as visualized in Fig.~\ref{fig:fig2}(a) for a 5QL-thick Bi$_2$Se$_3$ film under fixed electric field $\boldsymbol{E} = E \hat{x}$, corresponding to electron momentum predominantly in the $-x$ direction ($\phi_{\boldsymbol{k}} = \pi$). In absence of gate voltage, the scattering probability is strongly suppressed across all angles when $\theta_M$ is small, indicating efficient suppression of scattering regardless of the outgoing electron angle $\phi_{\boldsymbol{k}'}$. Increasing $\theta_M$ enhances the scattering amplitude, particularly at $\phi_{\boldsymbol{k}'} \approx \pi$, signaling enhanced resistivity. 

With this understanding, we next investigate the effect of magnetic impurity orientation on the longitudinal resistivity $\rho_{xx}$ in ultrathin magnetically doped TI films. This analysis is carried out within the framework of semi-classical Boltzmann transport theory, incorporating a modified relaxation time approximation~\cite{kohn1957quantum, vyborny2009semiclassical, sabzalipour2020two} to capture the anisotropic characteristics of electron–impurity scattering. The anisotropy originates from the spin-dependent nature of the scattering potential, which is inherently influenced by the tilt angle of the magnetic moments. While mechanisms such as anomalous velocity, arising from the Berry curvature, and side-jump effects, linked to spin-orbit-induced position shifts, contribute to transverse conductivity, they do not affect the longitudinal channel~\cite{sinitsyn2007anomalous, sabzalipour2019anomalous}. These effects are independent of impurity scattering and therefore excluded from our analysis, which focuses exclusively on scattering-driven contributions to the longitudinal charge current. 

Figure~\ref{fig:fig2}(b) presents the calculated longitudinal resistivity as a function of the magnetization tilt angle $\theta_M$ for TI films with thicknesses ranging from 2 to 5 QL. In our calculations, we consider the system in the dilute doping regime and the Fermi level is chosen at the upper bound of the single-band regime to isolate the contribution of the lowest conduction subband. The results reveal a non-monotonic angular dependence of $\rho_{xx}$. As $\theta_M$ increases from 0 (out-of-plane) toward $\pi/2$ (in-plane), the resistivity initially decreases, reaching a minimum at an intermediate angle that depends on the film thickness. Beyond this point, further tilting of the magnetization leads to an increase in resistivity. This behavior reflects the complex interplay between magnetization direction, spin texture, and hybridization gap. The observed minimum in $\rho_{xx}$ arises from the suppression of scattering when the magnetization is optimally aligned to minimize spin-flip scattering, while the subsequent increase at larger angles signals the dominance of in-plane magnetization effects that enhance scattering in specific momentum channels.

\subsection{Gate-controlled scattering and transport}
Electrostatic gating is known to be a potent tool for tuning the carrier density and modifying the band structure in quantum materials. In magnetically doped ultrathin TI systems, the interplay between surface states and band structure with electric field gives rise to a gate-tunable transport regime, enabling external control over electronic conduction. The effect of external gating is incorporated into the Hamiltonian [Eq.~\eqref{hamil_momen}] as an additional term $V_g \sigma_0 \otimes \tau_z$, where $V_g$ denotes the gate potential. 
Considering that the system itself already includes a built-in potential $V_{\mathrm{SIA}}$, 
the total potential in the system upon applying an external gate voltage is expressed as the sum of the structural inversion asymmetry and the gate contributions, i.e., $V_{total} = V_{SIA} + V_g$.

\begin{figure}[t]
    \centering
    \includegraphics[width=0.85\linewidth]{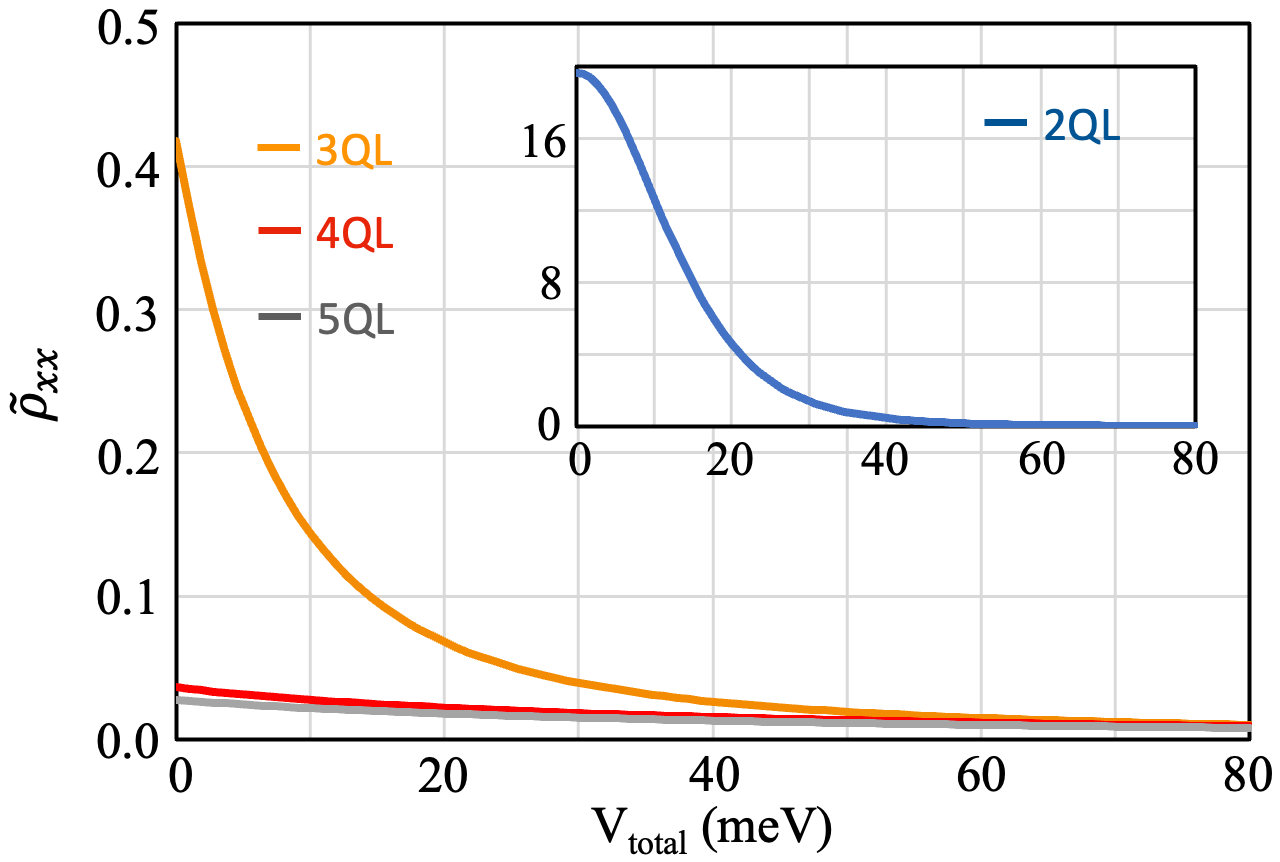}
    \caption{Longitudinal resistivity $\rho_{xx}$ as a function of total voltage for magnetically doped Bi$_2$Se$_3$ films with thicknesses of 2-5QL for a fixed magnetization angle $\theta_M = 0$. Increasing the voltage modifies the band structure and reduces spin-dependent scattering, resulting in a monotonic decrease in resistivity.}
    \label{fig:fig3}
\end{figure}

Here, we investigate the influence of gate voltage on the longitudinal resistivity $\rho_{xx}$ of magnetically doped Bi$_2$Se$_3$ films with thicknesses ranging from 2 to 5QL. The calculations are performed for a fixed magnetization direction $\theta_M = 0$ and for the Fermi level located at the upper bound of the single-band regime, as previously. Figure~\ref{fig:fig3} illustrates the computed variation of $\rho_{xx}$ as a function of the applied voltage. Applying a gate voltage modifies the band structure by shifting the relative position of the surface states, thereby affecting the available electronic states and the strength of spin-dependent scattering. As the gate voltage is increased, the effective scattering between electronic states is reduced.

\subsection{Combined control via magnetization and gating}
Having established that the orientation of magnetic impurities and the application of gate voltage can modulate the resistive response of topological surface states, we now examine how electrostatic gating, in conjunction with magnetization control, enables further tunability of scattering dynamics and charge transport. Experimentally, such control management is feasible, as the magnetization angle can be tuned by external field and probed using Hall sensors, the sample can be back-gated, and resistance measured via lock-in techniques~\cite{kawamura2018topological}.

\begin{figure}[t]
    \centering
    \includegraphics[width=0.82\linewidth]{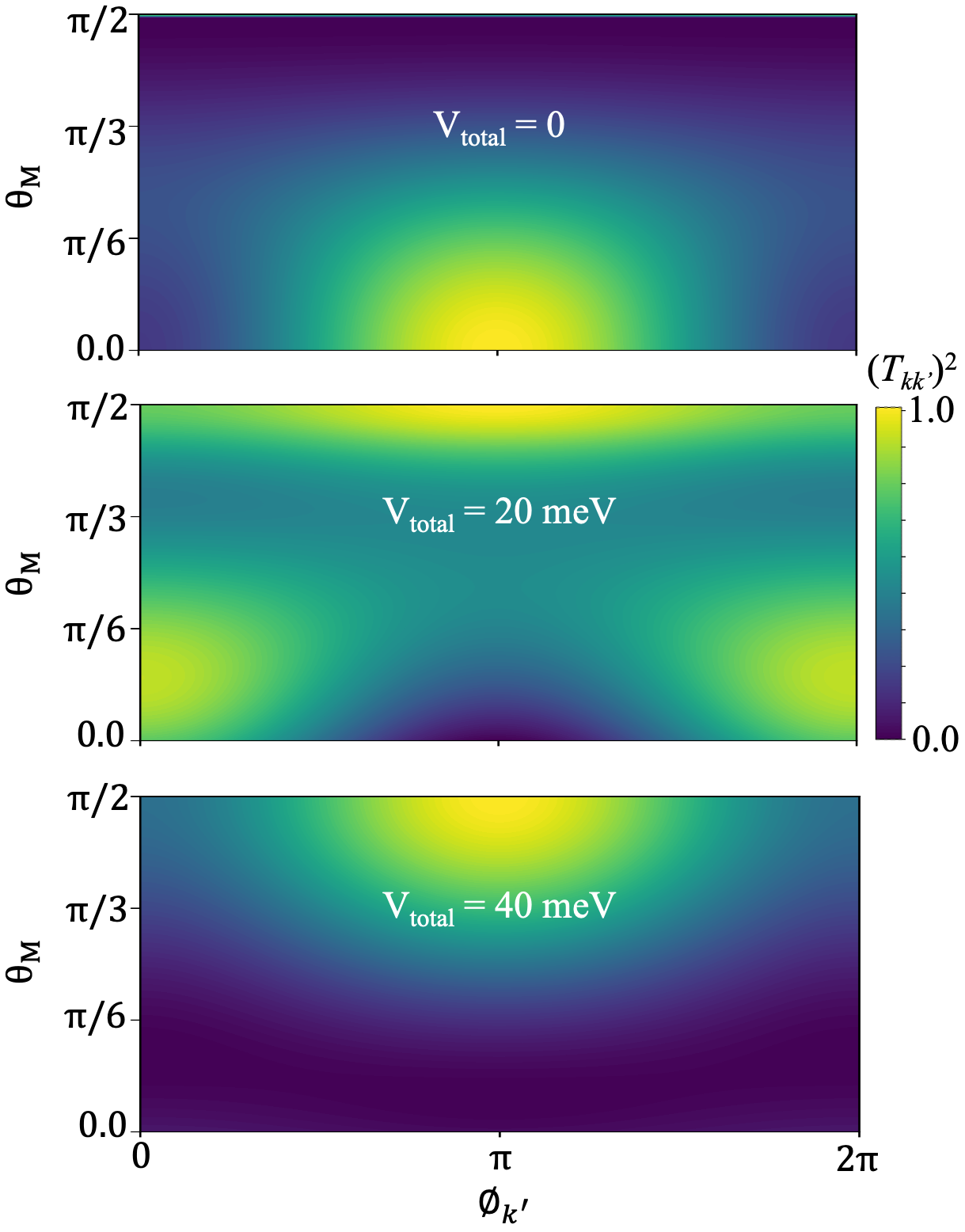}
    \caption{Gate-tunable scattering dynamics in a magnetically-doped 5QL-thick Bi$_2$Se$_3$ film. Angular scattering probability as a function of magnetization tilt angle $\theta_M$ and scattering angle $\phi_{\boldsymbol{k'}}$ for different total gate voltages, V$_{total}$ = V$_{SIA}$ + V$_g$ = 0, 20~meV and 40~meV. The scattering profiles reveal significant suppression of backscattering with applying gate voltage and magnetization tilt.}
    \label{fig:fig4}
\end{figure}

To explore the combined impact of gating and magnetization, we calculate the scattering probability as a function of the magnetization tilt angle $\theta_{M}$ and the scattering angle $\phi_{\boldsymbol{k'}}$ for a 5QL-thick magnetically doped Bi$_2$Se$_3$ film under distinct total voltages: V$_{total}$ = V$_{SIA}$ + V$_g$ = 0, 20~meV and 40~meV(see Fig.~\ref{fig:fig4}). These results reveal that both applying V$_{g}$ and tilting the magnetization control the probability of scattering, with near-complete suppression observed with some gating values and in a range of $\theta_M$ (for small tilt, $\theta_M < \pi/6$). %It should be noted that this behavior is observed primarily at small tilt angles. 

\begin{figure}[b]
    \centering
    \includegraphics[width=0.88\linewidth]{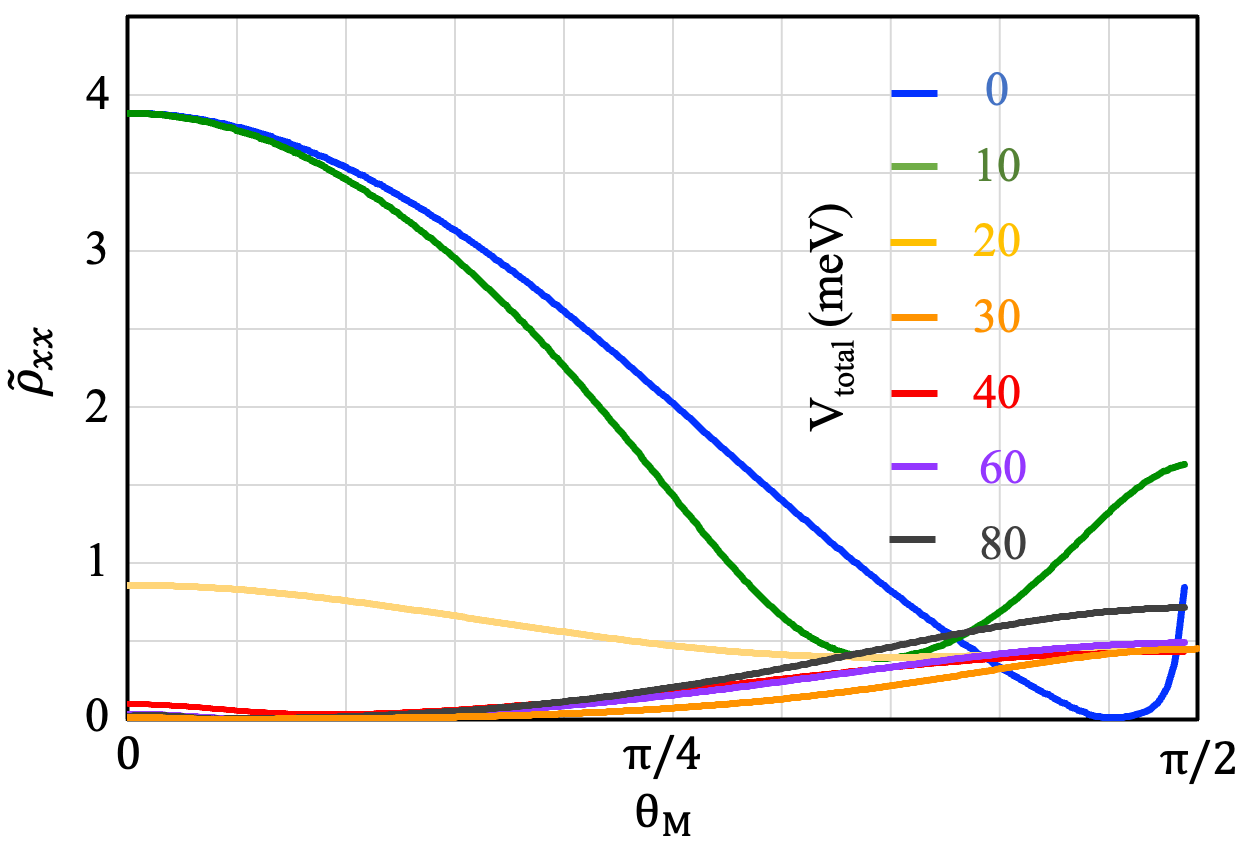}
    \caption{Longitudinal resistivity $\rho_{xx}$ as a function of $\theta_M$ for different V$_{total}$, demonstrating that combined gating and magnetization control can drastically reduce resistivity in the system.}
    \label{fig:fig5}
\end{figure}

To quantify the influence of these changes on transport properties, we compute the longitudinal resistivity (shown in Fig.~\ref{fig:fig5}) as a function of $\theta_{M}$ for different total voltages. The results confirm that the scattering is strongly suppressed with applying V$_{g}$, leading to significant reduction of resistivity of charge transport. Notably, as shown in Fig.~\ref{fig:fig5}, the resistivity in a 5QL-thick Bi$_2$Se$_3$ film entirely vanishes at sufficiently large voltage of V$_{total}\ge30$~meV for a range of magnetization tilt angles ($\theta_M < \pi/6$). This suppression of resistivity is a direct consequence of the suppression of the scattering rate, despite the finite density of magnetic impurities. It should be noted that, although our discussion thus far was focused on a 5QL sample, similar behavior is observed for other film thicknesses in the ultrathin regime, highlighting the generality of this effect. These findings demonstrate that by fine-tuning the gate voltage in conjunction with the magnetization angle, it is possible to completely suppress magnetic impurity-induced scattering, thereby achieving dissipationless transport in magnetically-doped TI thin films. %This suppression is largely independent of the hybridization gap, magnetic exchange coupling, and even the spatial arrangement of impurities, offering a promising route for designing tunable, low-dissipation electronic devices.

\section{Real-space calculations}\label{validation real-space}
To provide a closer relevance to realistic systems and to validate and complement the results obtained from the momentum-space effective Hamiltonian, we perform additional transport calculations based on a real-space tight-binding model. Namely, while the low-energy effective Hamiltonian provides a reliable description of the surface states of TIs, it inherently neglects several critical factors, including the influence of bulk states on surface transport, the dependence of the hybridization gap on magnetization orientation, and potential localization effects induced by in-plane magnetization components. These effects can play a significant role in ultrathin TI films, especially in regimes where bulk-surface coupling and disorder are non-negligible. To incorporate these complexities, we construct a real-space lattice Hamiltonian that captures both bulk and surface degrees of freedom, and compute the longitudinal conductance using the Landauer-B\"uttiker formalism, as detailed in the following.

\subsection{Hamiltonian and transport formalism}
To model the electronic transport properties of the system in real space, we employ a tight-binding Hamiltonian defined on a cubic lattice. The Hamiltonian can be expressed as~\cite{shafiei2022axion,chu2011surface}:
\begin{equation}\label{real_hamiltonian}
        H_{R} = \sum_{i} c_{i}^{\dagger} E_{\text{on}} c_{i} + \sum_{i, \boldsymbol{n}_i} \left( c_{i}^{\dagger} T_{\boldsymbol{n}_i} c_{i + \boldsymbol{n}_i} + \text{H.c.} \right),
\end{equation}
where \( c_i^\dagger \) (\( c_i \)) creates (annihilates) an electron at site \( i \), and \( \boldsymbol{n}_i = \hat{x}, \hat{y}, \hat{z} \) denotes the unit lattice vectors. The onsite energy term is given by \( E_{\text{on}} = (E_0 - 2\sum_{\boldsymbol{n}_i}B_{\boldsymbol{n}_i}) \sigma_z\otimes\sigma_0 \), and the hopping matrices are defined as \( T_{\boldsymbol{n}_i} = B_{\boldsymbol{n}_i} \sigma_z \otimes \sigma_0 - i \frac{A_{\boldsymbol{n}_i}}{2} \sigma_x \otimes \boldsymbol{\sigma} \cdot \boldsymbol{n}_i \). Here, \( E_0 \) denotes the bulk band gap, while \( A_{\boldsymbol{n}_i} \) and \( B_{\boldsymbol{n}_i} \) characterize the spin-orbit coupling and hopping strength in the respective directions. To parameterize this Hamiltonian, we adopt values obtained by fitting to DFT calculations for Bi$_2$Se$_3$~\cite{liu2010model}: \( A_{\parallel} = 0.5~\text{eV} \), \( A_z = 0.44~\text{eV} \), \( B_{\parallel} = B_z = 0.25~\text{eV} \), and \( E_0 = 0.28~\text{eV} \). A more detailed derivation and discussion of the model can be found in Ref.~\cite{shafiei2022axion}.

The interaction between the surface electrons and localized magnetic moments is introduced through an exchange term,
$H_{\text{ex}} = M_i \sigma_i \otimes \sigma_0 \, (i = x, y, z)$, where \( M_i \) denotes the Zeeman-type magnetic exchange field induced by magnetic impurities, resulting in a magnetic gap $\Delta_M = 2M_z$ in the system. In addition, the influence of electrostatic gating is incorporated via the potential term \( V_{total} \sigma_0 \otimes \sigma_z \), which enables tuning of the surface state degeneracy and the transport characteristics. %It is important to note that, in this model, the system is assumed to possess structural inversion symmetry. Therefore, the gate voltage in this context should be interpreted as an effective potential that includes contributions both from the external gate voltage and from the structural inversion asymmetry in the momentum-space model. 

To calculate the longitudinal conductance, we utilize the Landauer-B\"uttiker approach~\cite{datta1997electronic}. In this framework, the total conductance at zero temperature is given by $G = \frac{e^2}{h} \sum_n T_n(E_F)$, where \( T_n(E_F) \) represents the transmission probability of the \( n \)-th conducting channel at the Fermi energy \( E_F \). For multi-terminal configurations, the transmission between lead \( i \) and lead \( j \) is computed as $T_{ij} = \text{Tr} \left[ \Gamma_i G_{ij} \Gamma_j G_{ij}^\dagger \right]$, where \( \Gamma_i \) is the coupling matrix between the system and lead \( i \), and \( G_{ij} \) denotes the retarded Green's function between the respective lead interfaces.
Finally, the resistance is evaluated via $R_{ij} = \frac{1}{G_{ij}} - \frac{1}{N}$, where \( N \) denotes the number of propagating modes in the leads.

\subsection{Transport calculations}
\begin{figure}[t]
    \centering
    \includegraphics[width=0.9\linewidth]{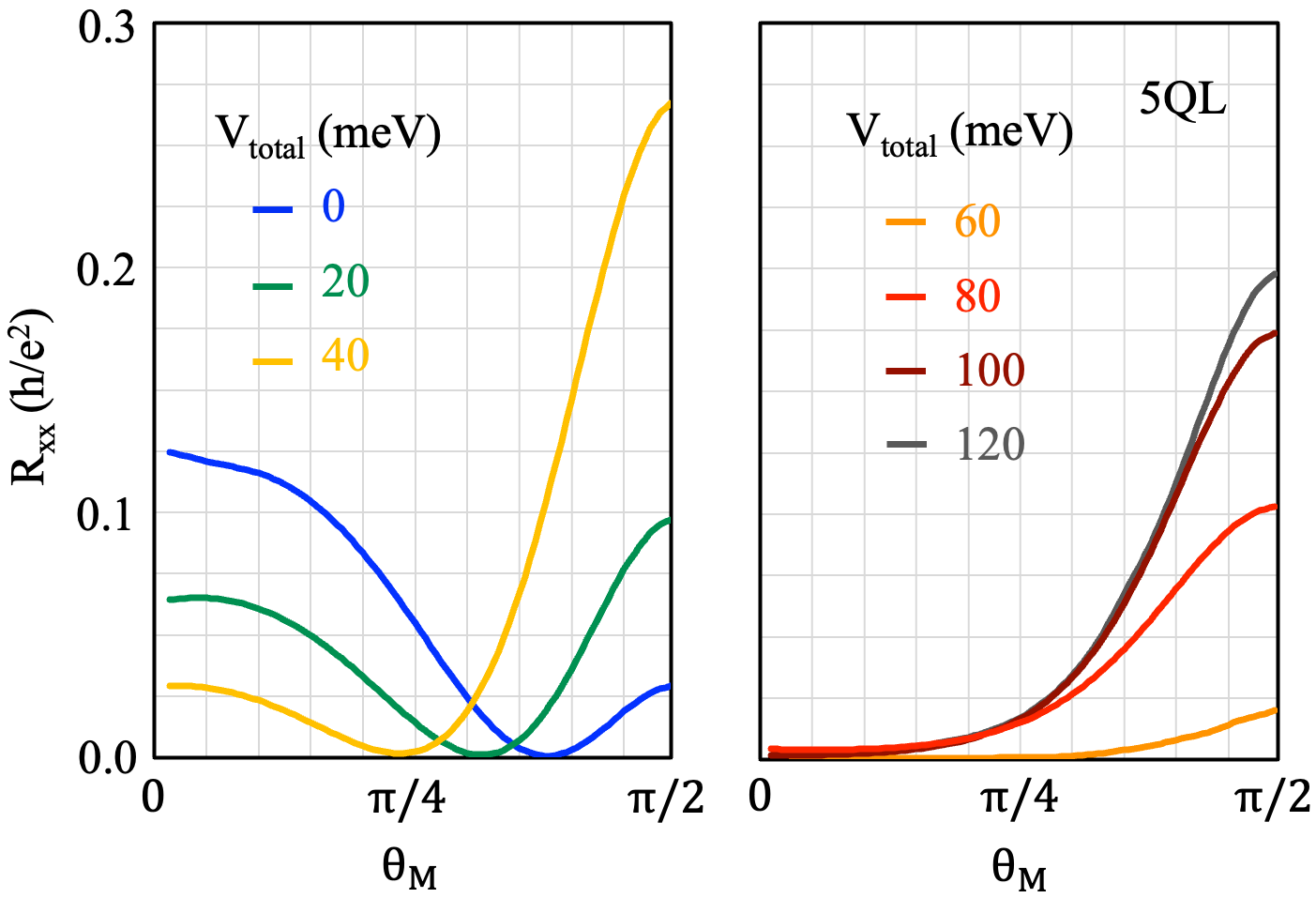}
    \caption{Calculated longitudinal resistance of a magnetically-doped 5QL Bi$_2$Se$_3$ film as a function of the magnetic impurity orientation angle \( \theta_M \) for different total voltages \( V_{total} \). Notably, the interplay between magnetic impurity orientation and electrostatic gating enables precise control over electronic transport, facilitating a transition to a dissipationless regime under optimal conditions.}
    \label{fig:fig6}
\end{figure}

Figure~\ref{fig:fig6} presents the calculated resistance of a 5QL thick magnetically doped TI film as a function of the magnetic impurity orientation angle \( \theta_M \) for various values of voltage ($V_{total}$). These results clearly demonstrate that longitudinal resistance is strongly dependent on \( \theta_M \), and can be minimized for specific angles. More importantly, the combination of proper magnetic impurity alignment and electrostatic gating can drive the system into a dissipationless transport regime. In these calculations, we set $\Delta_t$ = 40~meV, $\Delta_b$ = 5~meV, $\Delta_M$ = 15~meV and the Fermi energy is chosen to be the highest allowable value within the single-band regime, as in our previous considerations.

\begin{figure}[t]
    \centering
    \includegraphics[width=\linewidth]{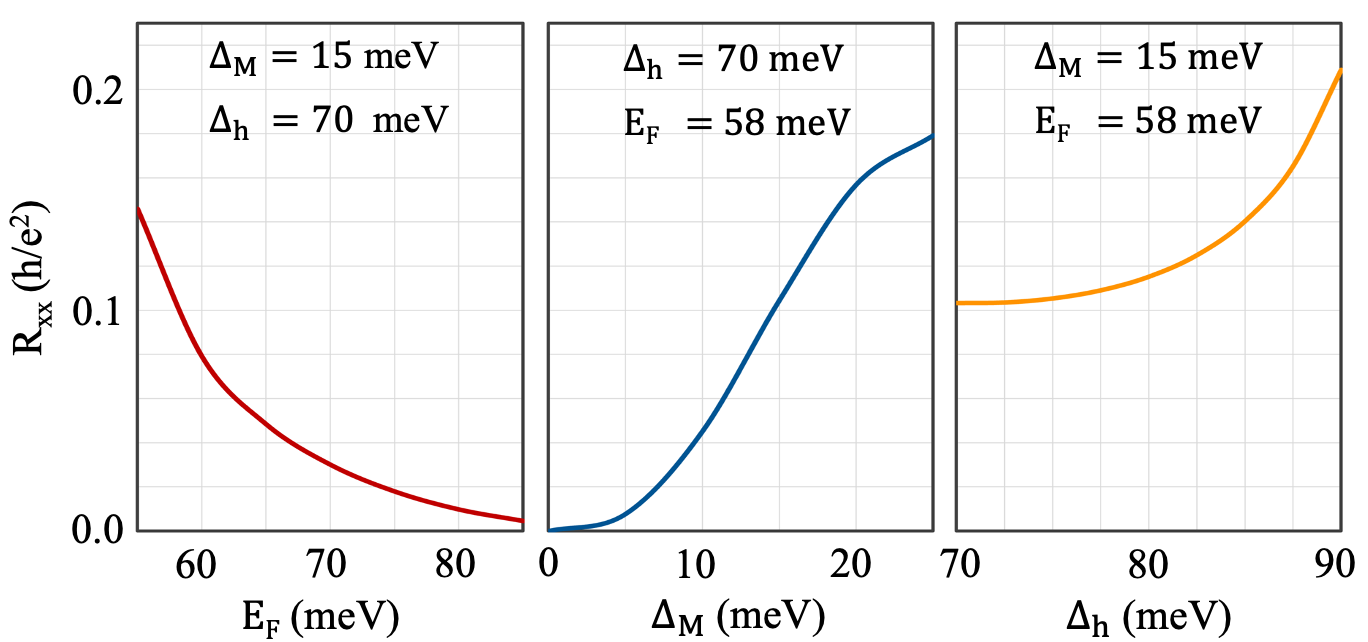}
    \caption{Longitudinal resistance of the magnetically doped ultrathin TI film as a function of Fermi energy, magnetic impurity concentration, and hybridization gap. The results highlight the distinct roles of these parameters in tuning the resistance within the single-band regime.}
    \label{fig:fig7}
\end{figure}

A comparison between the momentum- and real-space calculations shows their qualitative consistency across various parameter ranges, reinforcing the robustness of the observed phenomena. However, the real-space model reveals additional insights, particularly concerning the impact of in-plane magnetization. When \( \theta_M \) approaches \( \pi/2 \), the in-plane component of magnetization becomes dominant, leading to significant modifications in the transport behavior. This arises from a non-linear momentum-dependent term in the surface-state Hamiltonian, which distorts the Dirac cone and introduces anisotropic backscattering even in the presence of nonmagnetic impurities. Specifically, the in-plane magnetization can shift and tilt the Dirac cone, moving its center from \( (0,0) \) to \( (M_y, 0)/\hbar v_F \), thereby violating Lorentz invariance and enabling spin-flip backscattering between states \( \ket{k} \) and \( \ket{-k} \), as the spin orientations are no longer orthogonal~\cite{shafiei2025planar,zheng2020origin,imai2021spin}.

Our real-space quantum transport calculations confirm that despite the complexities introduced by bulk–surface coupling and magnetization-induced distortions, dissipationless transport states can still be achieved through careful tuning of the magnetization direction and external gating. This highlights the feasibility of practical device implementations based on topological insulator thin films with engineered magnetic configurations.

To further illustrate the individual impact of key parameters on the system resistance, we performed calculations of the longitudinal resistance by varying each relevant parameter independently, namely the Fermi energy, hybridization gap, and magnetic impurity concentration. Note that the hybridization gap is not a freely tunable parameter in real-space systems. Here, we controlled this parameter indirectly by applying external strain, as detailed in Ref.~\cite{shafiei2022controlling}. Although strain can, in general, modify the overall band structure of the system, its dominant effect in this context is on the hybridization gap. Therefore, this method has been employed as an effective means to modulate the hybridization gap in our analysis. The results of these calculations are summarized in Fig.~\ref{fig:fig7}. One sees that increasing the Fermi energy (while ensuring that it remains within the single-band regime) leads to a reduction in the system resistance. In contrast, increasing either the concentration of magnetic impurities or the hybridization gap results in a pronounced enhancement of the resistance.
\begin{figure}[t]
    \centering
    \includegraphics[width=0.9\linewidth]{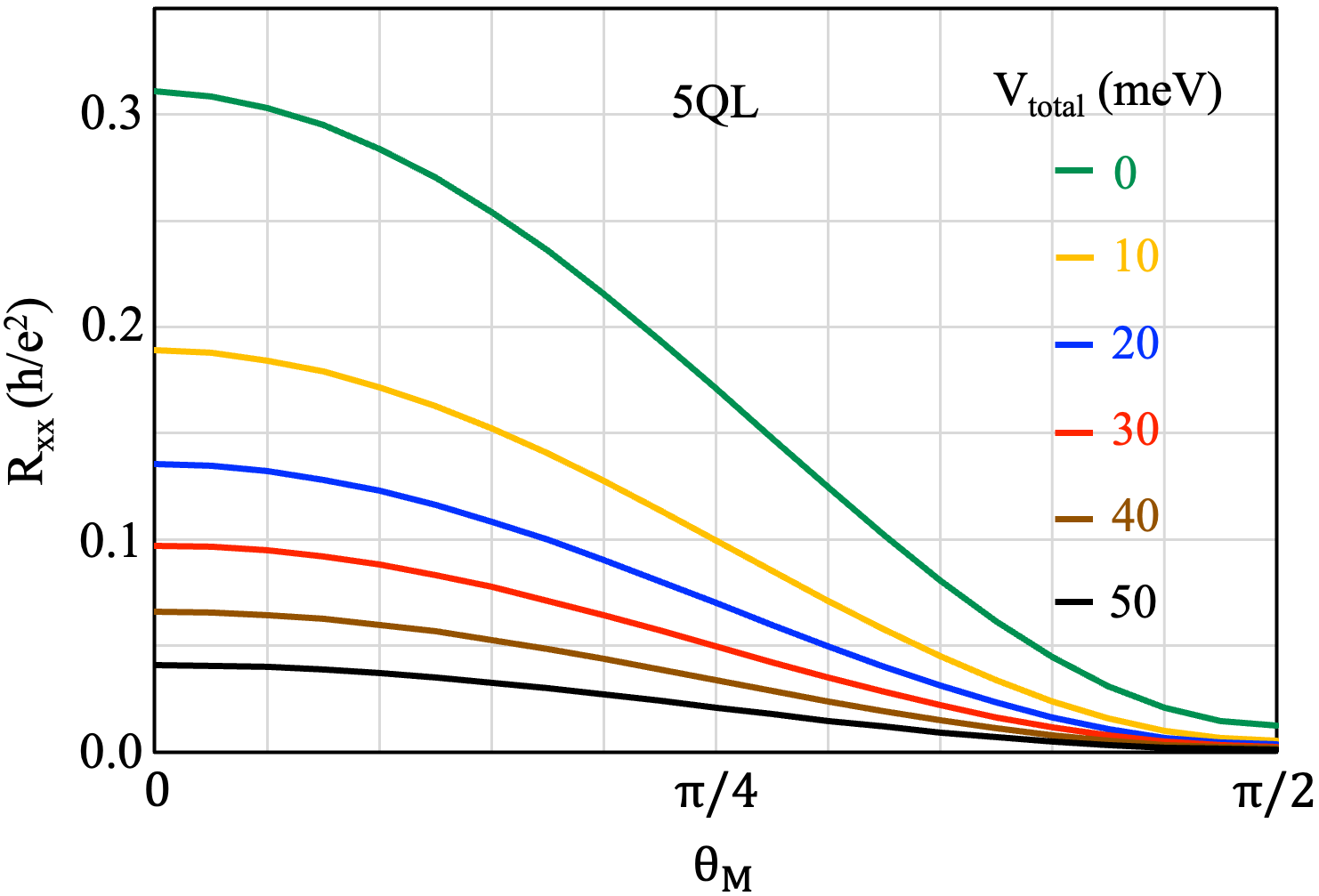}
    \caption{Calculated longitudinal resistance of a magnetically-doped 5QL TI film in the two-band regime, where both surface bands intersect the Fermi level (E$_F$ = 65 meV). Despite the emergence of additional intraband scattering channels compared to the single-band regime, the system retains high tunability.}
    \label{fig:fig8}
\end{figure}

Throughout the paper our focus has been on the single-band regime, where only one surface band crosses the Fermi level, for the already given reasons. For completeness, we have also performed transport calculations in the two-band regime for a 5QL magnetic TI and fixed E$_F$ = 65 meV (as shown in Fig.~\ref{fig:fig8}). In this accessible regime, the presence of both surface bands at the Fermi energy introduces additional interband scattering channels, leading to qualitative differences in transport behavior and the absence of a nonmonotonic dependence of the resistance on the key parameters. In fact, the presence of an additional conduction channel appears to mitigate the influence of the in-plane component of magnetic impurities on scattering due to the interband transition possibilities, resulting in a more uniform resistance response as a function of magnetization tilting. Nonetheless, our results show that even under these conditions, the system's transport properties remain highly tunable. Specifically, we demonstrate that tilting of the magnetization vector and/or applying electrostatic gating can still effectively suppress dissipation by reducing scattering between the relevant electronic states. These findings underscore the robustness of magnetization- and gate-controlled transport engineering in topological insulator thin films, extending its applicability beyond the idealized single-band scenario.

\section{Conclusions}\label{conclusions}
In this work, we have systematically investigated the role of magnetic impurity orientation and electrostatic gating in modulating the longitudinal transport properties of ultrathin magnetically-doped topological insulator (TI) films. Employing both momentum-space scattering analysis and real-space quantum transport calculations, we demonstrated that tilting the magnetization vector can induce significant changes in the electron scattering, thereby offering a powerful control knob over the electron transport. 

Our results reveal that the interplay between magnetic anisotropy and surface-state spin texture leads to a rich transport landscape, where resistance can be minimized - or even fully suppressed - by optimizing the magnetization direction. In particular, we identified the conditions under which dissipationless surface states emerge, facilitated by appropriate alignment of magnetic moments and presence of an external gate potential. These findings reveal a pathway towards robust, low-power spintronic devices using TI thin films with pre-engineered intrinsic magnetic textures.

Our focus was on the single-band regime, where we have shown a particularly successful, broad and precise control over the transport properties. In the two-band case, where both surface bands cross the Fermi level, the added scattering channels do hamper some of the strategies to reduce resistance, but the system remains highly tunable, and dissipation can still be strongly suppressed through magnetization tilting and applied gating - underscoring the robustness and generality of our transport control strategy. Altogether, our work advances the fundamental understanding of transport behavior in magnetic topological systems and provides practical design principles for the development of tunable, energy-efficient devices based on magnetic TIs.

\section*{Acknowledgments}
The authors thank Farhad Fazileh for useful discussions. This research was supported by the Research Foundation-Flanders (FWO-Vlaanderen), the Special Research Funds (BOF) of the University of Antwerp, and the FWO-FNRS EoS-ShapeME project.

\bibliography{bibliography}

\begin{thebibliography}{46}%
\makeatletter
\providecommand \@ifxundefined [1]{%
 \@ifx{#1\undefined}
}%
\providecommand \@ifnum [1]{%
 \ifnum #1\expandafter \@firstoftwo
 \else \expandafter \@secondoftwo
 \fi
}%
\providecommand \@ifx [1]{%
 \ifx #1\expandafter \@firstoftwo
 \else \expandafter \@secondoftwo
 \fi
}%
\providecommand \natexlab [1]{#1}%
\providecommand \enquote  [1]{``#1''}%
\providecommand \bibnamefont  [1]{#1}%
\providecommand \bibfnamefont [1]{#1}%
\providecommand \citenamefont [1]{#1}%
\providecommand \href@noop [0]{\@secondoftwo}%
\providecommand \href [0]{\begingroup \@sanitize@url \@href}%
\providecommand \@href[1]{\@@startlink{#1}\@@href}%
\providecommand \@@href[1]{\endgroup#1\@@endlink}%
\providecommand \@sanitize@url [0]{\catcode `\\12\catcode `\$12\catcode
  `\&12\catcode `\#12\catcode `\^12\catcode `\_12\catcode `\%12\relax}%
\providecommand \@@startlink[1]{}%
\providecommand \@@endlink[0]{}%
\providecommand \url  [0]{\begingroup\@sanitize@url \@url }%
\providecommand \@url [1]{\endgroup\@href {#1}{\urlprefix }}%
\providecommand \urlprefix  [0]{URL }%
\providecommand \Eprint [0]{\href }%
\providecommand \doibase [0]{https://doi.org/}%
\providecommand \selectlanguage [0]{\@gobble}%
\providecommand \bibinfo  [0]{\@secondoftwo}%
\providecommand \bibfield  [0]{\@secondoftwo}%
\providecommand \translation [1]{[#1]}%
\providecommand \BibitemOpen [0]{}%
\providecommand \bibitemStop [0]{}%
\providecommand \bibitemNoStop [0]{.\EOS\space}%
\providecommand \EOS [0]{\spacefactor3000\relax}%
\providecommand \BibitemShut  [1]{\csname bibitem#1\endcsname}%
\let\auto@bib@innerbib\@empty
%</preamble>
\bibitem [{\citenamefont {Tokura}\ \emph {et~al.}(2019)\citenamefont {Tokura},
  \citenamefont {Yasuda},\ and\ \citenamefont
  {Tsukazaki}}]{tokura2019magnetic}%
  \BibitemOpen
  \bibfield  {author} {\bibinfo {author} {\bibfnamefont {Y.}~\bibnamefont
  {Tokura}}, \bibinfo {author} {\bibfnamefont {K.}~\bibnamefont {Yasuda}},\
  and\ \bibinfo {author} {\bibfnamefont {A.}~\bibnamefont {Tsukazaki}},\
  }\bibfield  {title} {\bibinfo {title} {Magnetic topological insulators},\
  }\href@noop {} {\bibfield  {journal} {\bibinfo  {journal} {Nature Reviews
  Physics}\ }\textbf {\bibinfo {volume} {1}},\ \bibinfo {pages} {126} (\bibinfo
  {year} {2019})}\BibitemShut {NoStop}%
\bibitem [{\citenamefont {Fan}\ \emph {et~al.}(2016)\citenamefont {Fan},
  \citenamefont {Kou}, \citenamefont {Upadhyaya}, \citenamefont {Shao},
  \citenamefont {Pan}, \citenamefont {Lang}, \citenamefont {Che}, \citenamefont
  {Tang}, \citenamefont {Montazeri}, \citenamefont {Murata} \emph
  {et~al.}}]{fan2016electric}%
  \BibitemOpen
  \bibfield  {author} {\bibinfo {author} {\bibfnamefont {Y.}~\bibnamefont
  {Fan}}, \bibinfo {author} {\bibfnamefont {X.}~\bibnamefont {Kou}}, \bibinfo
  {author} {\bibfnamefont {P.}~\bibnamefont {Upadhyaya}}, \bibinfo {author}
  {\bibfnamefont {Q.}~\bibnamefont {Shao}}, \bibinfo {author} {\bibfnamefont
  {L.}~\bibnamefont {Pan}}, \bibinfo {author} {\bibfnamefont {M.}~\bibnamefont
  {Lang}}, \bibinfo {author} {\bibfnamefont {X.}~\bibnamefont {Che}}, \bibinfo
  {author} {\bibfnamefont {J.}~\bibnamefont {Tang}}, \bibinfo {author}
  {\bibfnamefont {M.}~\bibnamefont {Montazeri}}, \bibinfo {author}
  {\bibfnamefont {K.}~\bibnamefont {Murata}}, \emph {et~al.},\ }\bibfield
  {title} {\bibinfo {title} {Electric-field control of spin--orbit torque in a
  magnetically doped topological insulator},\ }\href@noop {} {\bibfield
  {journal} {\bibinfo  {journal} {Nature Nanotechnology}\ }\textbf {\bibinfo
  {volume} {11}},\ \bibinfo {pages} {352} (\bibinfo {year} {2016})}\BibitemShut
  {NoStop}%
\bibitem [{\citenamefont {Che}\ \emph {et~al.}(2020)\citenamefont {Che},
  \citenamefont {Pan}, \citenamefont {Vareskic}, \citenamefont {Zou},
  \citenamefont {Pan}, \citenamefont {Zhang}, \citenamefont {Yin},
  \citenamefont {Wu}, \citenamefont {Shao}, \citenamefont {Deng} \emph
  {et~al.}}]{che2020strongly}%
  \BibitemOpen
  \bibfield  {author} {\bibinfo {author} {\bibfnamefont {X.}~\bibnamefont
  {Che}}, \bibinfo {author} {\bibfnamefont {Q.}~\bibnamefont {Pan}}, \bibinfo
  {author} {\bibfnamefont {B.}~\bibnamefont {Vareskic}}, \bibinfo {author}
  {\bibfnamefont {J.}~\bibnamefont {Zou}}, \bibinfo {author} {\bibfnamefont
  {L.}~\bibnamefont {Pan}}, \bibinfo {author} {\bibfnamefont {P.}~\bibnamefont
  {Zhang}}, \bibinfo {author} {\bibfnamefont {G.}~\bibnamefont {Yin}}, \bibinfo
  {author} {\bibfnamefont {H.}~\bibnamefont {Wu}}, \bibinfo {author}
  {\bibfnamefont {Q.}~\bibnamefont {Shao}}, \bibinfo {author} {\bibfnamefont
  {P.}~\bibnamefont {Deng}}, \emph {et~al.},\ }\bibfield  {title} {\bibinfo
  {title} {Strongly surface state carrier-dependent spin--orbit torque in
  magnetic topological insulators},\ }\href@noop {} {\bibfield  {journal}
  {\bibinfo  {journal} {Advanced Materials}\ }\textbf {\bibinfo {volume}
  {32}},\ \bibinfo {pages} {1907661} (\bibinfo {year} {2020})}\BibitemShut
  {NoStop}%
\bibitem [{\citenamefont {Wang}\ \emph {et~al.}(2021)\citenamefont {Wang},
  \citenamefont {Ge}, \citenamefont {Li}, \citenamefont {Liu}, \citenamefont
  {Xu},\ and\ \citenamefont {Wang}}]{wang2021intrinsic}%
  \BibitemOpen
  \bibfield  {author} {\bibinfo {author} {\bibfnamefont {P.}~\bibnamefont
  {Wang}}, \bibinfo {author} {\bibfnamefont {J.}~\bibnamefont {Ge}}, \bibinfo
  {author} {\bibfnamefont {J.}~\bibnamefont {Li}}, \bibinfo {author}
  {\bibfnamefont {Y.}~\bibnamefont {Liu}}, \bibinfo {author} {\bibfnamefont
  {Y.}~\bibnamefont {Xu}},\ and\ \bibinfo {author} {\bibfnamefont
  {J.}~\bibnamefont {Wang}},\ }\bibfield  {title} {\bibinfo {title} {Intrinsic
  magnetic topological insulators},\ }\href@noop {} {\bibfield  {journal}
  {\bibinfo  {journal} {The Innovation}\ }\textbf {\bibinfo {volume} {2}}
  (\bibinfo {year} {2021})}\BibitemShut {NoStop}%
\bibitem [{\citenamefont {Bhattacharyya}\ \emph {et~al.}(2021)\citenamefont
  {Bhattacharyya}, \citenamefont {Mandal}, \citenamefont {Banerjee},
  \citenamefont {Barman}, \citenamefont {Das}, \citenamefont {Mondal},
  \citenamefont {Sengupta},\ and\ \citenamefont
  {Deb}}]{bhattacharyya2021recent}%
  \BibitemOpen
  \bibfield  {author} {\bibinfo {author} {\bibfnamefont {S.}~\bibnamefont
  {Bhattacharyya}}, \bibinfo {author} {\bibfnamefont {P.}~\bibnamefont
  {Mandal}}, \bibinfo {author} {\bibfnamefont {T.}~\bibnamefont {Banerjee}},
  \bibinfo {author} {\bibfnamefont {A.}~\bibnamefont {Barman}}, \bibinfo
  {author} {\bibfnamefont {S.}~\bibnamefont {Das}}, \bibinfo {author}
  {\bibfnamefont {M.}~\bibnamefont {Mondal}}, \bibinfo {author} {\bibfnamefont
  {K.}~\bibnamefont {Sengupta}},\ and\ \bibinfo {author} {\bibfnamefont
  {D.}~\bibnamefont {Deb}},\ }\bibfield  {title} {\bibinfo {title} {Recent
  progress in proximity coupling of magnetism to topological insulators},\
  }\href@noop {} {\bibfield  {journal} {\bibinfo  {journal} {Advanced
  Materials}\ }\textbf {\bibinfo {volume} {33}},\ \bibinfo {pages} {2007795}
  (\bibinfo {year} {2021})}\BibitemShut {NoStop}%
\bibitem [{\citenamefont {Chang}\ \emph {et~al.}(2013)\citenamefont {Chang},
  \citenamefont {Zhang}, \citenamefont {Feng}, \citenamefont {Shen},
  \citenamefont {Zhang}, \citenamefont {Guo}, \citenamefont {Li}, \citenamefont
  {Ou}, \citenamefont {Wei}, \citenamefont {Wang} \emph
  {et~al.}}]{chang2013experimental}%
  \BibitemOpen
  \bibfield  {author} {\bibinfo {author} {\bibfnamefont {C.-Z.}\ \bibnamefont
  {Chang}}, \bibinfo {author} {\bibfnamefont {J.}~\bibnamefont {Zhang}},
  \bibinfo {author} {\bibfnamefont {X.}~\bibnamefont {Feng}}, \bibinfo {author}
  {\bibfnamefont {J.}~\bibnamefont {Shen}}, \bibinfo {author} {\bibfnamefont
  {Z.}~\bibnamefont {Zhang}}, \bibinfo {author} {\bibfnamefont
  {M.}~\bibnamefont {Guo}}, \bibinfo {author} {\bibfnamefont {K.}~\bibnamefont
  {Li}}, \bibinfo {author} {\bibfnamefont {Y.}~\bibnamefont {Ou}}, \bibinfo
  {author} {\bibfnamefont {P.}~\bibnamefont {Wei}}, \bibinfo {author}
  {\bibfnamefont {L.-L.}\ \bibnamefont {Wang}}, \emph {et~al.},\ }\bibfield
  {title} {\bibinfo {title} {Experimental observation of the quantum anomalous
  hall effect in a magnetic topological insulator},\ }\href@noop {} {\bibfield
  {journal} {\bibinfo  {journal} {Science}\ }\textbf {\bibinfo {volume}
  {340}},\ \bibinfo {pages} {167} (\bibinfo {year} {2013})}\BibitemShut
  {NoStop}%
\bibitem [{\citenamefont {Fischer}\ \emph {et~al.}(2016)\citenamefont
  {Fischer}, \citenamefont {Vaezi}, \citenamefont {Manchon},\ and\
  \citenamefont {Kim}}]{fischer2016spin}%
  \BibitemOpen
  \bibfield  {author} {\bibinfo {author} {\bibfnamefont {M.~H.}\ \bibnamefont
  {Fischer}}, \bibinfo {author} {\bibfnamefont {A.}~\bibnamefont {Vaezi}},
  \bibinfo {author} {\bibfnamefont {A.}~\bibnamefont {Manchon}},\ and\ \bibinfo
  {author} {\bibfnamefont {E.-A.}\ \bibnamefont {Kim}},\ }\bibfield  {title}
  {\bibinfo {title} {Spin-torque generation in topological insulator based
  heterostructures},\ }\href@noop {} {\bibfield  {journal} {\bibinfo  {journal}
  {Physical Review B}\ }\textbf {\bibinfo {volume} {93}},\ \bibinfo {pages}
  {125303} (\bibinfo {year} {2016})}\BibitemShut {NoStop}%
\bibitem [{\citenamefont {Zhang}\ and\ \citenamefont
  {Fert}(2016)}]{zhang2016conversion}%
  \BibitemOpen
  \bibfield  {author} {\bibinfo {author} {\bibfnamefont {S.}~\bibnamefont
  {Zhang}}\ and\ \bibinfo {author} {\bibfnamefont {A.}~\bibnamefont {Fert}},\
  }\bibfield  {title} {\bibinfo {title} {Conversion between spin and charge
  currents with topological insulators},\ }\href@noop {} {\bibfield  {journal}
  {\bibinfo  {journal} {Physical Review B}\ }\textbf {\bibinfo {volume} {94}},\
  \bibinfo {pages} {184423} (\bibinfo {year} {2016})}\BibitemShut {NoStop}%
\bibitem [{\citenamefont {{\v{S}}mejkal}\ \emph {et~al.}(2018)\citenamefont
  {{\v{S}}mejkal}, \citenamefont {Mokrousov}, \citenamefont {Yan},\ and\
  \citenamefont {MacDonald}}]{vsmejkal2018topological}%
  \BibitemOpen
  \bibfield  {author} {\bibinfo {author} {\bibfnamefont {L.}~\bibnamefont
  {{\v{S}}mejkal}}, \bibinfo {author} {\bibfnamefont {Y.}~\bibnamefont
  {Mokrousov}}, \bibinfo {author} {\bibfnamefont {B.}~\bibnamefont {Yan}},\
  and\ \bibinfo {author} {\bibfnamefont {A.~H.}\ \bibnamefont {MacDonald}},\
  }\bibfield  {title} {\bibinfo {title} {Topological antiferromagnetic
  spintronics},\ }\href@noop {} {\bibfield  {journal} {\bibinfo  {journal}
  {Nature Physics}\ }\textbf {\bibinfo {volume} {14}},\ \bibinfo {pages} {242}
  (\bibinfo {year} {2018})}\BibitemShut {NoStop}%
\bibitem [{\citenamefont {Flatt{\'e}}(2017)}]{flatte2017voltage}%
  \BibitemOpen
  \bibfield  {author} {\bibinfo {author} {\bibfnamefont {M.~E.}\ \bibnamefont
  {Flatt{\'e}}},\ }\bibfield  {title} {\bibinfo {title} {Voltage-driven
  magnetization control in topological insulator/magnetic insulator
  heterostructures},\ }\href@noop {} {\bibfield  {journal} {\bibinfo  {journal}
  {AIP Advances}\ }\textbf {\bibinfo {volume} {7}} (\bibinfo {year}
  {2017})}\BibitemShut {NoStop}%
\bibitem [{\citenamefont {Wu}\ \emph {et~al.}(2014)\citenamefont {Wu},
  \citenamefont {Liu},\ and\ \citenamefont {Liu}}]{wu2014topological}%
  \BibitemOpen
  \bibfield  {author} {\bibinfo {author} {\bibfnamefont {J.}~\bibnamefont
  {Wu}}, \bibinfo {author} {\bibfnamefont {J.}~\bibnamefont {Liu}},\ and\
  \bibinfo {author} {\bibfnamefont {X.-J.}\ \bibnamefont {Liu}},\ }\bibfield
  {title} {\bibinfo {title} {Topological spin texture in a quantum anomalous
  hall insulator},\ }\href@noop {} {\bibfield  {journal} {\bibinfo  {journal}
  {Physical Review Letters}\ }\textbf {\bibinfo {volume} {113}},\ \bibinfo
  {pages} {136403} (\bibinfo {year} {2014})}\BibitemShut {NoStop}%
\bibitem [{\citenamefont {He}\ \emph {et~al.}(2019)\citenamefont {He},
  \citenamefont {Sun},\ and\ \citenamefont {He}}]{he2019topological}%
  \BibitemOpen
  \bibfield  {author} {\bibinfo {author} {\bibfnamefont {M.}~\bibnamefont
  {He}}, \bibinfo {author} {\bibfnamefont {H.}~\bibnamefont {Sun}},\ and\
  \bibinfo {author} {\bibfnamefont {Q.~L.}\ \bibnamefont {He}},\ }\bibfield
  {title} {\bibinfo {title} {Topological insulator: Spintronics and quantum
  computations},\ }\href@noop {} {\bibfield  {journal} {\bibinfo  {journal}
  {Frontiers of Physics}\ }\textbf {\bibinfo {volume} {14}},\ \bibinfo {pages}
  {1} (\bibinfo {year} {2019})}\BibitemShut {NoStop}%
\bibitem [{\citenamefont {Fan}\ and\ \citenamefont
  {Wang}(2016)}]{fan2016spintronics}%
  \BibitemOpen
  \bibfield  {author} {\bibinfo {author} {\bibfnamefont {Y.}~\bibnamefont
  {Fan}}\ and\ \bibinfo {author} {\bibfnamefont {K.~L.}\ \bibnamefont {Wang}},\
  }\bibfield  {title} {\bibinfo {title} {Spintronics based on topological
  insulators},\ }in\ \href@noop {} {\emph {\bibinfo {booktitle} {Spin}}},\
  Vol.~\bibinfo {volume} {6}\ (\bibinfo {organization} {World Scientific},\
  \bibinfo {year} {2016})\ p.\ \bibinfo {pages} {1640001}\BibitemShut {NoStop}%
\bibitem [{\citenamefont {Yu}\ \emph {et~al.}(2010)\citenamefont {Yu},
  \citenamefont {Zhang}, \citenamefont {Zhang}, \citenamefont {Zhang},
  \citenamefont {Dai},\ and\ \citenamefont {Fang}}]{yu2010quantized}%
  \BibitemOpen
  \bibfield  {author} {\bibinfo {author} {\bibfnamefont {R.}~\bibnamefont
  {Yu}}, \bibinfo {author} {\bibfnamefont {W.}~\bibnamefont {Zhang}}, \bibinfo
  {author} {\bibfnamefont {H.-J.}\ \bibnamefont {Zhang}}, \bibinfo {author}
  {\bibfnamefont {S.-C.}\ \bibnamefont {Zhang}}, \bibinfo {author}
  {\bibfnamefont {X.}~\bibnamefont {Dai}},\ and\ \bibinfo {author}
  {\bibfnamefont {Z.}~\bibnamefont {Fang}},\ }\bibfield  {title} {\bibinfo
  {title} {Quantized anomalous hall effect in magnetic topological
  insulators},\ }\href@noop {} {\bibfield  {journal} {\bibinfo  {journal}
  {Science}\ }\textbf {\bibinfo {volume} {329}},\ \bibinfo {pages} {61}
  (\bibinfo {year} {2010})}\BibitemShut {NoStop}%
\bibitem [{\citenamefont {Li}\ \emph {et~al.}(2014)\citenamefont {Li},
  \citenamefont {Van ‘T~Erve}, \citenamefont {Robinson}, \citenamefont {Liu},
  \citenamefont {Li},\ and\ \citenamefont {Jonker}}]{li2014electrical}%
  \BibitemOpen
  \bibfield  {author} {\bibinfo {author} {\bibfnamefont {C.}~\bibnamefont
  {Li}}, \bibinfo {author} {\bibfnamefont {O.}~\bibnamefont {Van ‘T~Erve}},
  \bibinfo {author} {\bibfnamefont {J.}~\bibnamefont {Robinson}}, \bibinfo
  {author} {\bibfnamefont {Y.}~\bibnamefont {Liu}}, \bibinfo {author}
  {\bibfnamefont {L.}~\bibnamefont {Li}},\ and\ \bibinfo {author}
  {\bibfnamefont {B.}~\bibnamefont {Jonker}},\ }\bibfield  {title} {\bibinfo
  {title} {Electrical detection of charge-current-induced spin polarization due
  to spin-momentum locking in {Bi2Se3}},\ }\href@noop {} {\bibfield  {journal}
  {\bibinfo  {journal} {Nature Nanotechnology}\ }\textbf {\bibinfo {volume}
  {9}},\ \bibinfo {pages} {218} (\bibinfo {year} {2014})}\BibitemShut {NoStop}%
\bibitem [{\citenamefont {Mellnik}\ \emph {et~al.}(2014)\citenamefont
  {Mellnik}, \citenamefont {Lee}, \citenamefont {Richardella}, \citenamefont
  {Grab}, \citenamefont {Mintun}, \citenamefont {Fischer}, \citenamefont
  {Vaezi}, \citenamefont {Manchon}, \citenamefont {Kim}, \citenamefont
  {Samarth} \emph {et~al.}}]{mellnik2014spin}%
  \BibitemOpen
  \bibfield  {author} {\bibinfo {author} {\bibfnamefont {A.}~\bibnamefont
  {Mellnik}}, \bibinfo {author} {\bibfnamefont {J.}~\bibnamefont {Lee}},
  \bibinfo {author} {\bibfnamefont {A.}~\bibnamefont {Richardella}}, \bibinfo
  {author} {\bibfnamefont {J.}~\bibnamefont {Grab}}, \bibinfo {author}
  {\bibfnamefont {P.}~\bibnamefont {Mintun}}, \bibinfo {author} {\bibfnamefont
  {M.~H.}\ \bibnamefont {Fischer}}, \bibinfo {author} {\bibfnamefont
  {A.}~\bibnamefont {Vaezi}}, \bibinfo {author} {\bibfnamefont
  {A.}~\bibnamefont {Manchon}}, \bibinfo {author} {\bibfnamefont {E.-A.}\
  \bibnamefont {Kim}}, \bibinfo {author} {\bibfnamefont {N.}~\bibnamefont
  {Samarth}}, \emph {et~al.},\ }\bibfield  {title} {\bibinfo {title}
  {Spin-transfer torque generated by a topological insulator},\ }\href@noop {}
  {\bibfield  {journal} {\bibinfo  {journal} {Nature}\ }\textbf {\bibinfo
  {volume} {511}},\ \bibinfo {pages} {449} (\bibinfo {year}
  {2014})}\BibitemShut {NoStop}%
\bibitem [{\citenamefont {Wang}\ \emph {et~al.}(2023)\citenamefont {Wang},
  \citenamefont {Wu}, \citenamefont {Zhang}, \citenamefont {Liu}, \citenamefont
  {Chen}, \citenamefont {Pandey}, \citenamefont {Yin}, \citenamefont {Wei},
  \citenamefont {Lei}, \citenamefont {Shi} \emph {et~al.}}]{wang2023room}%
  \BibitemOpen
  \bibfield  {author} {\bibinfo {author} {\bibfnamefont {H.}~\bibnamefont
  {Wang}}, \bibinfo {author} {\bibfnamefont {H.}~\bibnamefont {Wu}}, \bibinfo
  {author} {\bibfnamefont {J.}~\bibnamefont {Zhang}}, \bibinfo {author}
  {\bibfnamefont {Y.}~\bibnamefont {Liu}}, \bibinfo {author} {\bibfnamefont
  {D.}~\bibnamefont {Chen}}, \bibinfo {author} {\bibfnamefont {C.}~\bibnamefont
  {Pandey}}, \bibinfo {author} {\bibfnamefont {J.}~\bibnamefont {Yin}},
  \bibinfo {author} {\bibfnamefont {D.}~\bibnamefont {Wei}}, \bibinfo {author}
  {\bibfnamefont {N.}~\bibnamefont {Lei}}, \bibinfo {author} {\bibfnamefont
  {S.}~\bibnamefont {Shi}}, \emph {et~al.},\ }\bibfield  {title} {\bibinfo
  {title} {Room temperature energy-efficient spin-orbit torque switching in
  two-dimensional van der waals {Fe3GeTe2} induced by topological insulators},\
  }\href@noop {} {\bibfield  {journal} {\bibinfo  {journal} {Nature
  Communications}\ }\textbf {\bibinfo {volume} {14}},\ \bibinfo {pages} {5173}
  (\bibinfo {year} {2023})}\BibitemShut {NoStop}%
\bibitem [{\citenamefont {Binda}\ \emph {et~al.}(2023)\citenamefont {Binda},
  \citenamefont {Fedel}, \citenamefont {Alvarado-Guti{\'e}rrez}, \citenamefont
  {No{\"e}l},\ and\ \citenamefont {Gambardella}}]{binda2023large}%
  \BibitemOpen
  \bibfield  {author} {\bibinfo {author} {\bibfnamefont {F.}~\bibnamefont
  {Binda}}, \bibinfo {author} {\bibfnamefont {S.}~\bibnamefont {Fedel}},
  \bibinfo {author} {\bibfnamefont {S.~F.}\ \bibnamefont
  {Alvarado-Guti{\'e}rrez}}, \bibinfo {author} {\bibfnamefont {P.}~\bibnamefont
  {No{\"e}l}},\ and\ \bibinfo {author} {\bibfnamefont {P.}~\bibnamefont
  {Gambardella}},\ }\bibfield  {title} {\bibinfo {title} {Large and isotropic
  spin-orbit torques and spin hall magnetoresistance from the twin-free
  topological insulator {Bi0.9Sb0.1}},\ }\href@noop {} {\bibfield  {journal}
  {\bibinfo  {journal} {Advanced Materials}\ } (\bibinfo {year}
  {2023})}\BibitemShut {NoStop}%
\bibitem [{\citenamefont {Wu}\ \emph {et~al.}(2019)\citenamefont {Wu},
  \citenamefont {Zhang}, \citenamefont {Deng}, \citenamefont {Lan},
  \citenamefont {Pan}, \citenamefont {Razavi}, \citenamefont {Che},
  \citenamefont {Huang}, \citenamefont {Dai}, \citenamefont {Wong} \emph
  {et~al.}}]{wu2019room}%
  \BibitemOpen
  \bibfield  {author} {\bibinfo {author} {\bibfnamefont {H.}~\bibnamefont
  {Wu}}, \bibinfo {author} {\bibfnamefont {P.}~\bibnamefont {Zhang}}, \bibinfo
  {author} {\bibfnamefont {P.}~\bibnamefont {Deng}}, \bibinfo {author}
  {\bibfnamefont {Q.}~\bibnamefont {Lan}}, \bibinfo {author} {\bibfnamefont
  {Q.}~\bibnamefont {Pan}}, \bibinfo {author} {\bibfnamefont {S.~A.}\
  \bibnamefont {Razavi}}, \bibinfo {author} {\bibfnamefont {X.}~\bibnamefont
  {Che}}, \bibinfo {author} {\bibfnamefont {L.}~\bibnamefont {Huang}}, \bibinfo
  {author} {\bibfnamefont {B.}~\bibnamefont {Dai}}, \bibinfo {author}
  {\bibfnamefont {K.}~\bibnamefont {Wong}}, \emph {et~al.},\ }\bibfield
  {title} {\bibinfo {title} {Room-temperature spin-orbit torque from
  topological surface states},\ }\href@noop {} {\bibfield  {journal} {\bibinfo
  {journal} {Physical Review Letters}\ }\textbf {\bibinfo {volume} {123}},\
  \bibinfo {pages} {207205} (\bibinfo {year} {2019})}\BibitemShut {NoStop}%
\bibitem [{\citenamefont {Haazen}\ \emph {et~al.}(2012)\citenamefont {Haazen},
  \citenamefont {Lalo{\"e}}, \citenamefont {Nummy}, \citenamefont {Swagten},
  \citenamefont {Jarillo-Herrero}, \citenamefont {Heiman},\ and\ \citenamefont
  {Moodera}}]{haazen2012ferromagnetism}%
  \BibitemOpen
  \bibfield  {author} {\bibinfo {author} {\bibfnamefont {P.}~\bibnamefont
  {Haazen}}, \bibinfo {author} {\bibfnamefont {J.-B.}\ \bibnamefont
  {Lalo{\"e}}}, \bibinfo {author} {\bibfnamefont {T.}~\bibnamefont {Nummy}},
  \bibinfo {author} {\bibfnamefont {H.}~\bibnamefont {Swagten}}, \bibinfo
  {author} {\bibfnamefont {P.}~\bibnamefont {Jarillo-Herrero}}, \bibinfo
  {author} {\bibfnamefont {D.}~\bibnamefont {Heiman}},\ and\ \bibinfo {author}
  {\bibfnamefont {J.}~\bibnamefont {Moodera}},\ }\bibfield  {title} {\bibinfo
  {title} {Ferromagnetism in thin-film cr-doped topological insulator
  {Bi2Se3}},\ }\href@noop {} {\bibfield  {journal} {\bibinfo  {journal}
  {Applied Physics Letters}\ }\textbf {\bibinfo {volume} {100}} (\bibinfo
  {year} {2012})}\BibitemShut {NoStop}%
\bibitem [{\citenamefont {Kou}\ \emph {et~al.}(2013)\citenamefont {Kou},
  \citenamefont {Lang}, \citenamefont {Fan}, \citenamefont {Jiang},
  \citenamefont {Nie}, \citenamefont {Zhang}, \citenamefont {Jiang},
  \citenamefont {Wang}, \citenamefont {Yao}, \citenamefont {He} \emph
  {et~al.}}]{kou2013interplay}%
  \BibitemOpen
  \bibfield  {author} {\bibinfo {author} {\bibfnamefont {X.}~\bibnamefont
  {Kou}}, \bibinfo {author} {\bibfnamefont {M.}~\bibnamefont {Lang}}, \bibinfo
  {author} {\bibfnamefont {Y.}~\bibnamefont {Fan}}, \bibinfo {author}
  {\bibfnamefont {Y.}~\bibnamefont {Jiang}}, \bibinfo {author} {\bibfnamefont
  {T.}~\bibnamefont {Nie}}, \bibinfo {author} {\bibfnamefont {J.}~\bibnamefont
  {Zhang}}, \bibinfo {author} {\bibfnamefont {W.}~\bibnamefont {Jiang}},
  \bibinfo {author} {\bibfnamefont {Y.}~\bibnamefont {Wang}}, \bibinfo {author}
  {\bibfnamefont {Y.}~\bibnamefont {Yao}}, \bibinfo {author} {\bibfnamefont
  {L.}~\bibnamefont {He}}, \emph {et~al.},\ }\bibfield  {title} {\bibinfo
  {title} {Interplay between different magnetisms in cr-doped topological
  insulators},\ }\href@noop {} {\bibfield  {journal} {\bibinfo  {journal} {ACS
  nano}\ }\textbf {\bibinfo {volume} {7}},\ \bibinfo {pages} {9205} (\bibinfo
  {year} {2013})}\BibitemShut {NoStop}%
\bibitem [{\citenamefont {Pop}(2010)}]{pop2010energy}%
  \BibitemOpen
  \bibfield  {author} {\bibinfo {author} {\bibfnamefont {E.}~\bibnamefont
  {Pop}},\ }\bibfield  {title} {\bibinfo {title} {Energy dissipation and
  transport in nanoscale devices},\ }\href@noop {} {\bibfield  {journal}
  {\bibinfo  {journal} {Nano Research}\ }\textbf {\bibinfo {volume} {3}},\
  \bibinfo {pages} {147} (\bibinfo {year} {2010})}\BibitemShut {NoStop}%
\bibitem [{\citenamefont {Vassighi}\ and\ \citenamefont
  {Sachdev}(2006)}]{vassighi2006thermal}%
  \BibitemOpen
  \bibfield  {author} {\bibinfo {author} {\bibfnamefont {A.}~\bibnamefont
  {Vassighi}}\ and\ \bibinfo {author} {\bibfnamefont {M.}~\bibnamefont
  {Sachdev}},\ }\href@noop {} {\emph {\bibinfo {title} {Thermal and power
  management of integrated circuits}}}\ (\bibinfo  {publisher} {Springer
  Science \& Business Media},\ \bibinfo {year} {2006})\BibitemShut {NoStop}%
\bibitem [{\citenamefont {Yan}\ \emph {et~al.}(2024)\citenamefont {Yan},
  \citenamefont {Li}, \citenamefont {Jiang}, \citenamefont {Sun},\ and\
  \citenamefont {Xie}}]{yan2024rules}%
  \BibitemOpen
  \bibfield  {author} {\bibinfo {author} {\bibfnamefont {Q.}~\bibnamefont
  {Yan}}, \bibinfo {author} {\bibfnamefont {H.}~\bibnamefont {Li}}, \bibinfo
  {author} {\bibfnamefont {H.}~\bibnamefont {Jiang}}, \bibinfo {author}
  {\bibfnamefont {Q.-F.}\ \bibnamefont {Sun}},\ and\ \bibinfo {author}
  {\bibfnamefont {X.}~\bibnamefont {Xie}},\ }\bibfield  {title} {\bibinfo
  {title} {Rules for dissipationless topotronics},\ }\href@noop {} {\bibfield
  {journal} {\bibinfo  {journal} {Science Advances}\ }\textbf {\bibinfo
  {volume} {10}},\ \bibinfo {pages} {eado4756} (\bibinfo {year}
  {2024})}\BibitemShut {NoStop}%
\bibitem [{\citenamefont {Zhang}\ \emph {et~al.}(2009)\citenamefont {Zhang},
  \citenamefont {Liu}, \citenamefont {Qi}, \citenamefont {Dai}, \citenamefont
  {Fang},\ and\ \citenamefont {Zhang}}]{zhang2009topological}%
  \BibitemOpen
  \bibfield  {author} {\bibinfo {author} {\bibfnamefont {H.}~\bibnamefont
  {Zhang}}, \bibinfo {author} {\bibfnamefont {C.-X.}\ \bibnamefont {Liu}},
  \bibinfo {author} {\bibfnamefont {X.-L.}\ \bibnamefont {Qi}}, \bibinfo
  {author} {\bibfnamefont {X.}~\bibnamefont {Dai}}, \bibinfo {author}
  {\bibfnamefont {Z.}~\bibnamefont {Fang}},\ and\ \bibinfo {author}
  {\bibfnamefont {S.-C.}\ \bibnamefont {Zhang}},\ }\bibfield  {title} {\bibinfo
  {title} {Topological insulators in {Bi2Se3}, {Bi2Te3} and {Sb2Te3} with a
  single dirac cone on the surface},\ }\href@noop {} {\bibfield  {journal}
  {\bibinfo  {journal} {Nature Physics}\ }\textbf {\bibinfo {volume} {5}},\
  \bibinfo {pages} {438} (\bibinfo {year} {2009})}\BibitemShut {NoStop}%
\bibitem [{\citenamefont {Zhang}\ \emph {et~al.}(2010)\citenamefont {Zhang},
  \citenamefont {He}, \citenamefont {Chang}, \citenamefont {Song},
  \citenamefont {Wang}, \citenamefont {Chen}, \citenamefont {Jia},
  \citenamefont {Fang}, \citenamefont {Dai}, \citenamefont {Shan} \emph
  {et~al.}}]{zhang2010crossover}%
  \BibitemOpen
  \bibfield  {author} {\bibinfo {author} {\bibfnamefont {Y.}~\bibnamefont
  {Zhang}}, \bibinfo {author} {\bibfnamefont {K.}~\bibnamefont {He}}, \bibinfo
  {author} {\bibfnamefont {C.-Z.}\ \bibnamefont {Chang}}, \bibinfo {author}
  {\bibfnamefont {C.-L.}\ \bibnamefont {Song}}, \bibinfo {author}
  {\bibfnamefont {L.-L.}\ \bibnamefont {Wang}}, \bibinfo {author}
  {\bibfnamefont {X.}~\bibnamefont {Chen}}, \bibinfo {author} {\bibfnamefont
  {J.-F.}\ \bibnamefont {Jia}}, \bibinfo {author} {\bibfnamefont
  {Z.}~\bibnamefont {Fang}}, \bibinfo {author} {\bibfnamefont {X.}~\bibnamefont
  {Dai}}, \bibinfo {author} {\bibfnamefont {W.-Y.}\ \bibnamefont {Shan}}, \emph
  {et~al.},\ }\bibfield  {title} {\bibinfo {title} {Crossover of the
  three-dimensional topological insulator {Bi2Se3} to the two-dimensional
  limit},\ }\href@noop {} {\bibfield  {journal} {\bibinfo  {journal} {Nature
  Physics}\ }\textbf {\bibinfo {volume} {6}},\ \bibinfo {pages} {584} (\bibinfo
  {year} {2010})}\BibitemShut {NoStop}%
\bibitem [{\citenamefont {Shafiei}\ \emph {et~al.}(2024)\citenamefont
  {Shafiei}, \citenamefont {Fazileh}, \citenamefont {Peeters},\ and\
  \citenamefont {Milo{\v{s}}evi{\'c}}}]{shafiei2024tuning}%
  \BibitemOpen
  \bibfield  {author} {\bibinfo {author} {\bibfnamefont {M.}~\bibnamefont
  {Shafiei}}, \bibinfo {author} {\bibfnamefont {F.}~\bibnamefont {Fazileh}},
  \bibinfo {author} {\bibfnamefont {F.~M.}\ \bibnamefont {Peeters}},\ and\
  \bibinfo {author} {\bibfnamefont {M.~V.}\ \bibnamefont
  {Milo{\v{s}}evi{\'c}}},\ }\bibfield  {title} {\bibinfo {title} {Tuning the
  quantum phase transition of an ultrathin magnetic topological insulator},\
  }\href@noop {} {\bibfield  {journal} {\bibinfo  {journal} {Physical Review
  Materials}\ }\textbf {\bibinfo {volume} {8}},\ \bibinfo {pages} {074201}
  (\bibinfo {year} {2024})}\BibitemShut {NoStop}%
\bibitem [{\citenamefont {Shan}\ \emph {et~al.}(2010)\citenamefont {Shan},
  \citenamefont {Lu},\ and\ \citenamefont {Shen}}]{shan2010effective}%
  \BibitemOpen
  \bibfield  {author} {\bibinfo {author} {\bibfnamefont {W.-Y.}\ \bibnamefont
  {Shan}}, \bibinfo {author} {\bibfnamefont {H.-Z.}\ \bibnamefont {Lu}},\ and\
  \bibinfo {author} {\bibfnamefont {S.-Q.}\ \bibnamefont {Shen}},\ }\bibfield
  {title} {\bibinfo {title} {Effective continuous model for surface states and
  thin films of three-dimensional topological insulators},\ }\href@noop {}
  {\bibfield  {journal} {\bibinfo  {journal} {New Journal of Physics}\ }\textbf
  {\bibinfo {volume} {12}},\ \bibinfo {pages} {043048} (\bibinfo {year}
  {2010})}\BibitemShut {NoStop}%
\bibitem [{Note1()}]{Note1}%
  \BibitemOpen
  \bibinfo {note} {In Ref.~\cite {hou2020axion}, first-principles calculations
  have demonstrated that sandwiching a TI film between CrI$_3$ and
  MnBi$_2$Se$_4$ induces distinctly different exchange fields (approximately
  2.9~meV and 26.9~meV, respectively) on the top and bottom surfaces of the
  TI.}\BibitemShut {Stop}%
\bibitem [{\citenamefont {Li}\ \emph {et~al.}(2022)\citenamefont {Li},
  \citenamefont {Trang}, \citenamefont {Wu}, \citenamefont {Hwang},
  \citenamefont {Cortie}, \citenamefont {Medhekar}, \citenamefont {Mo},
  \citenamefont {Yang},\ and\ \citenamefont {Edmonds}}]{li2022large}%
  \BibitemOpen
  \bibfield  {author} {\bibinfo {author} {\bibfnamefont {Q.}~\bibnamefont
  {Li}}, \bibinfo {author} {\bibfnamefont {C.~X.}\ \bibnamefont {Trang}},
  \bibinfo {author} {\bibfnamefont {W.}~\bibnamefont {Wu}}, \bibinfo {author}
  {\bibfnamefont {J.}~\bibnamefont {Hwang}}, \bibinfo {author} {\bibfnamefont
  {D.}~\bibnamefont {Cortie}}, \bibinfo {author} {\bibfnamefont
  {N.}~\bibnamefont {Medhekar}}, \bibinfo {author} {\bibfnamefont {S.-K.}\
  \bibnamefont {Mo}}, \bibinfo {author} {\bibfnamefont {S.~A.}\ \bibnamefont
  {Yang}},\ and\ \bibinfo {author} {\bibfnamefont {M.~T.}\ \bibnamefont
  {Edmonds}},\ }\bibfield  {title} {\bibinfo {title} {Large magnetic gap in a
  designer ferromagnet--topological insulator--ferromagnet heterostructure},\
  }\href@noop {} {\bibfield  {journal} {\bibinfo  {journal} {Advanced
  Materials}\ }\textbf {\bibinfo {volume} {34}},\ \bibinfo {pages} {2107520}
  (\bibinfo {year} {2022})}\BibitemShut {NoStop}%
\bibitem [{\citenamefont {Eremeev}\ \emph {et~al.}(2018)\citenamefont
  {Eremeev}, \citenamefont {Otrokov},\ and\ \citenamefont
  {Chulkov}}]{eremeev2018new}%
  \BibitemOpen
  \bibfield  {author} {\bibinfo {author} {\bibfnamefont {S.~V.}\ \bibnamefont
  {Eremeev}}, \bibinfo {author} {\bibfnamefont {M.~M.}\ \bibnamefont
  {Otrokov}},\ and\ \bibinfo {author} {\bibfnamefont {E.~V.}\ \bibnamefont
  {Chulkov}},\ }\bibfield  {title} {\bibinfo {title} {New universal type of
  interface in the magnetic insulator/topological insulator heterostructures},\
  }\href@noop {} {\bibfield  {journal} {\bibinfo  {journal} {Nano letters}\
  }\textbf {\bibinfo {volume} {18}},\ \bibinfo {pages} {6521} (\bibinfo {year}
  {2018})}\BibitemShut {NoStop}%
\bibitem [{\citenamefont {Kohn}\ and\ \citenamefont
  {Luttinger}(1957)}]{kohn1957quantum}%
  \BibitemOpen
  \bibfield  {author} {\bibinfo {author} {\bibfnamefont {W.}~\bibnamefont
  {Kohn}}\ and\ \bibinfo {author} {\bibfnamefont {J.}~\bibnamefont
  {Luttinger}},\ }\bibfield  {title} {\bibinfo {title} {Quantum theory of
  electrical transport phenomena},\ }\href@noop {} {\bibfield  {journal}
  {\bibinfo  {journal} {Physical Review B}\ }\textbf {\bibinfo {volume}
  {108}},\ \bibinfo {pages} {590} (\bibinfo {year} {1957})}\BibitemShut
  {NoStop}%
\bibitem [{\citenamefont {V{\`y}born{\`y}}\ \emph {et~al.}(2009)\citenamefont
  {V{\`y}born{\`y}}, \citenamefont {Kovalev}, \citenamefont {Sinova},\ and\
  \citenamefont {Jungwirth}}]{vyborny2009semiclassical}%
  \BibitemOpen
  \bibfield  {author} {\bibinfo {author} {\bibfnamefont {K.}~\bibnamefont
  {V{\`y}born{\`y}}}, \bibinfo {author} {\bibfnamefont {A.~A.}\ \bibnamefont
  {Kovalev}}, \bibinfo {author} {\bibfnamefont {J.}~\bibnamefont {Sinova}},\
  and\ \bibinfo {author} {\bibfnamefont {T.}~\bibnamefont {Jungwirth}},\
  }\bibfield  {title} {\bibinfo {title} {Semiclassical framework for the
  calculation of transport anisotropies},\ }\href@noop {} {\bibfield  {journal}
  {\bibinfo  {journal} {Physical Review B}\ }\textbf {\bibinfo {volume} {79}},\
  \bibinfo {pages} {045427} (\bibinfo {year} {2009})}\BibitemShut {NoStop}%
\bibitem [{\citenamefont {Sabzalipour}\ \emph {et~al.}(2020)\citenamefont
  {Sabzalipour}, \citenamefont {Mir}, \citenamefont {Zarenia},\ and\
  \citenamefont {Partoens}}]{sabzalipour2020two}%
  \BibitemOpen
  \bibfield  {author} {\bibinfo {author} {\bibfnamefont {A.}~\bibnamefont
  {Sabzalipour}}, \bibinfo {author} {\bibfnamefont {M.}~\bibnamefont {Mir}},
  \bibinfo {author} {\bibfnamefont {M.}~\bibnamefont {Zarenia}},\ and\ \bibinfo
  {author} {\bibfnamefont {B.}~\bibnamefont {Partoens}},\ }\bibfield  {title}
  {\bibinfo {title} {Two distinctive regimes in the charge transport of a
  magnetic topological ultra thin film},\ }\href@noop {} {\bibfield  {journal}
  {\bibinfo  {journal} {New Journal of Physics}\ }\textbf {\bibinfo {volume}
  {22}},\ \bibinfo {pages} {123004} (\bibinfo {year} {2020})}\BibitemShut
  {NoStop}%
\bibitem [{\citenamefont {Sinitsyn}\ \emph {et~al.}(2007)\citenamefont
  {Sinitsyn}, \citenamefont {MacDonald}, \citenamefont {Jungwirth},
  \citenamefont {Dugaev},\ and\ \citenamefont
  {Sinova}}]{sinitsyn2007anomalous}%
  \BibitemOpen
  \bibfield  {author} {\bibinfo {author} {\bibfnamefont {N.}~\bibnamefont
  {Sinitsyn}}, \bibinfo {author} {\bibfnamefont {A.}~\bibnamefont {MacDonald}},
  \bibinfo {author} {\bibfnamefont {T.}~\bibnamefont {Jungwirth}}, \bibinfo
  {author} {\bibfnamefont {V.}~\bibnamefont {Dugaev}},\ and\ \bibinfo {author}
  {\bibfnamefont {J.}~\bibnamefont {Sinova}},\ }\bibfield  {title} {\bibinfo
  {title} {Anomalous hall effect in a two-dimensional dirac band: The link
  between the kubo-streda formula and the semiclassical boltzmann equation
  approach},\ }\href@noop {} {\bibfield  {journal} {\bibinfo  {journal}
  {Physical Review B}\ }\textbf {\bibinfo {volume} {75}},\ \bibinfo {pages}
  {045315} (\bibinfo {year} {2007})}\BibitemShut {NoStop}%
\bibitem [{\citenamefont {Sabzalipour}\ and\ \citenamefont
  {Partoens}(2019)}]{sabzalipour2019anomalous}%
  \BibitemOpen
  \bibfield  {author} {\bibinfo {author} {\bibfnamefont {A.}~\bibnamefont
  {Sabzalipour}}\ and\ \bibinfo {author} {\bibfnamefont {B.}~\bibnamefont
  {Partoens}},\ }\bibfield  {title} {\bibinfo {title} {Anomalous hall effect in
  magnetic topological insulators: Semiclassical framework},\ }\href@noop {}
  {\bibfield  {journal} {\bibinfo  {journal} {Physical Review B}\ }\textbf
  {\bibinfo {volume} {100}},\ \bibinfo {pages} {035419} (\bibinfo {year}
  {2019})}\BibitemShut {NoStop}%
\bibitem [{\citenamefont {Kawamura}\ \emph {et~al.}(2018)\citenamefont
  {Kawamura}, \citenamefont {Mogi}, \citenamefont {Yoshimi}, \citenamefont
  {Tsukazaki}, \citenamefont {Kozuka}, \citenamefont {Takahashi}, \citenamefont
  {Kawasaki},\ and\ \citenamefont {Tokura}}]{kawamura2018topological}%
  \BibitemOpen
  \bibfield  {author} {\bibinfo {author} {\bibfnamefont {M.}~\bibnamefont
  {Kawamura}}, \bibinfo {author} {\bibfnamefont {M.}~\bibnamefont {Mogi}},
  \bibinfo {author} {\bibfnamefont {R.}~\bibnamefont {Yoshimi}}, \bibinfo
  {author} {\bibfnamefont {A.}~\bibnamefont {Tsukazaki}}, \bibinfo {author}
  {\bibfnamefont {Y.}~\bibnamefont {Kozuka}}, \bibinfo {author} {\bibfnamefont
  {K.~S.}\ \bibnamefont {Takahashi}}, \bibinfo {author} {\bibfnamefont
  {M.}~\bibnamefont {Kawasaki}},\ and\ \bibinfo {author} {\bibfnamefont
  {Y.}~\bibnamefont {Tokura}},\ }\bibfield  {title} {\bibinfo {title}
  {Topological quantum phase transition in magnetic topological insulator upon
  magnetization rotation},\ }\href@noop {} {\bibfield  {journal} {\bibinfo
  {journal} {Physical Review B}\ }\textbf {\bibinfo {volume} {98}},\ \bibinfo
  {pages} {140404} (\bibinfo {year} {2018})}\BibitemShut {NoStop}%
\bibitem [{\citenamefont {Shafiei}\ \emph
  {et~al.}(2022{\natexlab{a}})\citenamefont {Shafiei}, \citenamefont {Fazileh},
  \citenamefont {Peeters},\ and\ \citenamefont
  {Milo{\v{s}}evi{\'c}}}]{shafiei2022axion}%
  \BibitemOpen
  \bibfield  {author} {\bibinfo {author} {\bibfnamefont {M.}~\bibnamefont
  {Shafiei}}, \bibinfo {author} {\bibfnamefont {F.}~\bibnamefont {Fazileh}},
  \bibinfo {author} {\bibfnamefont {F.~M.}\ \bibnamefont {Peeters}},\ and\
  \bibinfo {author} {\bibfnamefont {M.~V.}\ \bibnamefont
  {Milo{\v{s}}evi{\'c}}},\ }\bibfield  {title} {\bibinfo {title} {Axion
  insulator states in a topological insulator proximitized to magnetic
  insulators: A tight-binding characterization},\ }\href@noop {} {\bibfield
  {journal} {\bibinfo  {journal} {Physical Review Materials}\ }\textbf
  {\bibinfo {volume} {6}},\ \bibinfo {pages} {074205} (\bibinfo {year}
  {2022}{\natexlab{a}})}\BibitemShut {NoStop}%
\bibitem [{\citenamefont {Chu}\ \emph {et~al.}(2011)\citenamefont {Chu},
  \citenamefont {Shi},\ and\ \citenamefont {Shen}}]{chu2011surface}%
  \BibitemOpen
  \bibfield  {author} {\bibinfo {author} {\bibfnamefont {R.-L.}\ \bibnamefont
  {Chu}}, \bibinfo {author} {\bibfnamefont {J.}~\bibnamefont {Shi}},\ and\
  \bibinfo {author} {\bibfnamefont {S.-Q.}\ \bibnamefont {Shen}},\ }\bibfield
  {title} {\bibinfo {title} {Surface edge state and half-quantized hall
  conductance in topological insulators},\ }\href@noop {} {\bibfield  {journal}
  {\bibinfo  {journal} {Physical Review B}\ }\textbf {\bibinfo {volume} {84}},\
  \bibinfo {pages} {085312} (\bibinfo {year} {2011})}\BibitemShut {NoStop}%
\bibitem [{\citenamefont {Liu}\ \emph {et~al.}(2010)\citenamefont {Liu},
  \citenamefont {Qi}, \citenamefont {Zhang}, \citenamefont {Dai}, \citenamefont
  {Fang},\ and\ \citenamefont {Zhang}}]{liu2010model}%
  \BibitemOpen
  \bibfield  {author} {\bibinfo {author} {\bibfnamefont {C.-X.}\ \bibnamefont
  {Liu}}, \bibinfo {author} {\bibfnamefont {X.-L.}\ \bibnamefont {Qi}},
  \bibinfo {author} {\bibfnamefont {H.}~\bibnamefont {Zhang}}, \bibinfo
  {author} {\bibfnamefont {X.}~\bibnamefont {Dai}}, \bibinfo {author}
  {\bibfnamefont {Z.}~\bibnamefont {Fang}},\ and\ \bibinfo {author}
  {\bibfnamefont {S.-C.}\ \bibnamefont {Zhang}},\ }\bibfield  {title} {\bibinfo
  {title} {Model hamiltonian for topological insulators},\ }\href@noop {}
  {\bibfield  {journal} {\bibinfo  {journal} {Physical Review B}\ }\textbf
  {\bibinfo {volume} {82}},\ \bibinfo {pages} {045122} (\bibinfo {year}
  {2010})}\BibitemShut {NoStop}%
\bibitem [{\citenamefont {Datta}(1997)}]{datta1997electronic}%
  \BibitemOpen
  \bibfield  {author} {\bibinfo {author} {\bibfnamefont {S.}~\bibnamefont
  {Datta}},\ }\href@noop {} {\emph {\bibinfo {title} {Electronic transport in
  mesoscopic systems}}}\ (\bibinfo  {publisher} {Cambridge university press},\
  \bibinfo {year} {1997})\BibitemShut {NoStop}%
\bibitem [{\citenamefont {Shafiei}\ and\ \citenamefont
  {Milo{\v{s}}evi{\'c}}(2025)}]{shafiei2025planar}%
  \BibitemOpen
  \bibfield  {author} {\bibinfo {author} {\bibfnamefont {M.}~\bibnamefont
  {Shafiei}}\ and\ \bibinfo {author} {\bibfnamefont {M.~V.}\ \bibnamefont
  {Milo{\v{s}}evi{\'c}}},\ }\bibfield  {title} {\bibinfo {title} {Planar hall
  effect in ultrathin topological insulator films},\ }\href@noop {} {\bibfield
  {journal} {\bibinfo  {journal} {Physical Review B}\ }\textbf {\bibinfo
  {volume} {112}},\ \bibinfo {pages} {035401} (\bibinfo {year}
  {2025})}\BibitemShut {NoStop}%
\bibitem [{\citenamefont {Zheng}\ \emph {et~al.}(2020)\citenamefont {Zheng},
  \citenamefont {Duan}, \citenamefont {Wang}, \citenamefont {Li}, \citenamefont
  {Deng},\ and\ \citenamefont {Wang}}]{zheng2020origin}%
  \BibitemOpen
  \bibfield  {author} {\bibinfo {author} {\bibfnamefont {S.-H.}\ \bibnamefont
  {Zheng}}, \bibinfo {author} {\bibfnamefont {H.-J.}\ \bibnamefont {Duan}},
  \bibinfo {author} {\bibfnamefont {J.-K.}\ \bibnamefont {Wang}}, \bibinfo
  {author} {\bibfnamefont {J.-Y.}\ \bibnamefont {Li}}, \bibinfo {author}
  {\bibfnamefont {M.-X.}\ \bibnamefont {Deng}},\ and\ \bibinfo {author}
  {\bibfnamefont {R.-Q.}\ \bibnamefont {Wang}},\ }\bibfield  {title} {\bibinfo
  {title} {Origin of planar hall effect on the surface of topological
  insulators: Tilt of dirac cone by an in-plane magnetic field},\ }\href@noop
  {} {\bibfield  {journal} {\bibinfo  {journal} {Physical Review B}\ }\textbf
  {\bibinfo {volume} {101}},\ \bibinfo {pages} {041408} (\bibinfo {year}
  {2020})}\BibitemShut {NoStop}%
\bibitem [{\citenamefont {Imai}\ \emph {et~al.}(2021)\citenamefont {Imai},
  \citenamefont {Yamakage},\ and\ \citenamefont {Kohno}}]{imai2021spin}%
  \BibitemOpen
  \bibfield  {author} {\bibinfo {author} {\bibfnamefont {Y.}~\bibnamefont
  {Imai}}, \bibinfo {author} {\bibfnamefont {A.}~\bibnamefont {Yamakage}},\
  and\ \bibinfo {author} {\bibfnamefont {H.}~\bibnamefont {Kohno}},\ }\bibfield
   {title} {\bibinfo {title} {Spin-orbit torques and magnetotransport of
  two-dimensional dirac electrons without particle-hole symmetry},\ }\href@noop
  {} {\bibfield  {journal} {\bibinfo  {journal} {Physical Review B}\ }\textbf
  {\bibinfo {volume} {103}},\ \bibinfo {pages} {144416} (\bibinfo {year}
  {2021})}\BibitemShut {NoStop}%
\bibitem [{\citenamefont {Shafiei}\ \emph
  {et~al.}(2022{\natexlab{b}})\citenamefont {Shafiei}, \citenamefont {Fazileh},
  \citenamefont {Peeters},\ and\ \citenamefont
  {Milo{\v{s}}evi{\'c}}}]{shafiei2022controlling}%
  \BibitemOpen
  \bibfield  {author} {\bibinfo {author} {\bibfnamefont {M.}~\bibnamefont
  {Shafiei}}, \bibinfo {author} {\bibfnamefont {F.}~\bibnamefont {Fazileh}},
  \bibinfo {author} {\bibfnamefont {F.~M.}\ \bibnamefont {Peeters}},\ and\
  \bibinfo {author} {\bibfnamefont {M.~V.}\ \bibnamefont
  {Milo{\v{s}}evi{\'c}}},\ }\bibfield  {title} {\bibinfo {title} {Controlling
  the hybridization gap and transport in a thin-film topological insulator:
  Effect of strain, and electric and magnetic field},\ }\href@noop {}
  {\bibfield  {journal} {\bibinfo  {journal} {Physical Review B}\ }\textbf
  {\bibinfo {volume} {106}},\ \bibinfo {pages} {035119} (\bibinfo {year}
  {2022}{\natexlab{b}})}\BibitemShut {NoStop}%
\bibitem [{\citenamefont {Hou}\ \emph {et~al.}(2020)\citenamefont {Hou},
  \citenamefont {Kim},\ and\ \citenamefont {Wu}}]{hou2020axion}%
  \BibitemOpen
  \bibfield  {author} {\bibinfo {author} {\bibfnamefont {Y.}~\bibnamefont
  {Hou}}, \bibinfo {author} {\bibfnamefont {J.}~\bibnamefont {Kim}},\ and\
  \bibinfo {author} {\bibfnamefont {R.}~\bibnamefont {Wu}},\ }\bibfield
  {title} {\bibinfo {title} {Axion insulator state in ferromagnetically ordered
  {CrI3}/{Bi2Se3}/{MnBi2Se4} heterostructures},\ }\href@noop {} {\bibfield
  {journal} {\bibinfo  {journal} {Physical Review B}\ }\textbf {\bibinfo
  {volume} {101}},\ \bibinfo {pages} {121401} (\bibinfo {year}
  {2020})}\BibitemShut {NoStop}%
\end{thebibliography}%

\end{document}